\documentclass[prx,superscriptaddress,floatfix,showpacs,notitlepage,twocolumn]{revtex4-2}
\usepackage[latin9]{inputenc}
\usepackage{color}
\usepackage{textcomp}
\usepackage{amsmath}
\usepackage{amssymb}
\usepackage{bbold}
\usepackage{mathtools}
\usepackage{graphicx}
\usepackage{esint}
\usepackage{framed} 
\usepackage[dvipsnames]{xcolor}
\usepackage{hyperref}
\usepackage{enumerate}
\usepackage{enumitem}
\usepackage{moreenum}

\usepackage[caption=false]{subfig}
\hypersetup{
  linkcolor = black,
  citecolor  = blue,
  urlcolor = black,
  colorlinks = true,
}
\usepackage{dirtytalk}
\makeatletter
  \definecolor{color0}  {RGB}{12, 35, 64} 

\usepackage{amsfonts}
\usepackage{natbib}
\usepackage{svg}
\usepackage{float}
\usepackage{lipsum}
\usepackage[export]{adjustbox}
\usepackage{xcolor}
\hypersetup{    colorlinks,    linkcolor={blue!50!black},    citecolor={blue!50!black},    urlcolor={blue!80!black}}
\usepackage[all]{hypcap}
\usepackage{bbold}

\newcommand*{\bra}[1]{\ensuremath{\langle #1 \vert}}
\newcommand*{\ket}[1]{\ensuremath{\vert #1 \rangle}}

\newcommand*{\tr}[1]{\mathrm{tr}\left(#1\right)}
\newcommand*{\ptr}[2]{\mathrm{tr}_{#1}\left(#2\right)}

\newcommand{\mc}[1]{\mathcal{#1}}

\newcommand{\lr}[1]{\left( #1 \right)}
\newcommand{\mean}[1]{\langle#1\rangle}

\newcommand\mydots{\hbox to 1em{.\hss.\hss.}}

\begin{document}

\title{Continuous Coherent Quantum Feedback with Time Delays: Tensor Network Solution}
 \author{Kseniia Vodenkova}
 \affiliation{Institute for Theoretical Physics, University of Innsbruck, 6020 Innsbruck, Austria}
 \affiliation{Institute for Quantum Optics and Quantum Information of the Austrian Academy of Sciences, 6020 Innsbruck, Austria}
 \author{Hannes Pichler}
 \email{hannes.pichler@uibk.ac.at}
 \affiliation{Institute for Theoretical Physics, University of Innsbruck, 6020 Innsbruck, Austria}
 \affiliation{Institute for Quantum Optics and Quantum Information of the Austrian Academy of Sciences, 6020 Innsbruck, Austria}

\begin{abstract}
 In this paper we develop a novel method to solve problems involving quantum optical systems coupled to coherent quantum feedback loops  featuring time delays. Our method is based on exact mappings of such non-Markovian problems to equivalent Markovian driven dissipative quantum many-body problems. In this work we show that the resulting Markovian quantum many-body problems can be solved  (numerically) exactly and efficiently using tensor network methods for a series of paradigmatic examples, consisting of driven quantum systems coupled to waveguides at several distant points. In particular, we show that our method allows solving problems in so far inaccessible regimes, including problems with arbitrary long time delays and arbitrary numbers of excitations in the delay lines. We obtain solutions for the full real-time dynamics as well as the steady state in all these regimes. Finally, motivated by our results, we develop a novel mean-field approach, which allows us to find the solution semi-analytically and identify parameter regimes where this approximation is in excellent agreement with our exact tensor network results.
\end{abstract}

\maketitle
\section{Introduction}

Feedback is a cornerstone concept in modern technology, serving as the backbone for optimization and control in complex systems, where feedback loops take data from systems, process it, and adjust system parameters to achieve the desired outcome.
Quantum feedback refers to the situation when the system of interest is quantum mechanical in nature \cite{wisemanQuantumTheoryOptical1993,wisemanQuantumTheoryContinuous1994,wisemanQuantumMeasurementControl2010,dohertyQuantumFeedbackControl2000,zhangQuantumFeedbackTheory2017,kubanekPhotonbyphotonFeedbackControl2009,sayrinRealtimeQuantumFeedback2011,vijayStabilizingRabiOscillations2012,hiroseCoherentFeedbackControl2016,magriniRealtimeOptimalQuantum2021}. Here one can distinguish between two classes of feedback. In conventional, measurement-based quantum feedback, data is taken by projective or weak measurements, processed classically, and then used to adjust classical controls of the quantum system \cite{wisemanQuantumTheoryOptical1993,wisemanQuantumTheoryContinuous1994}. In contrast, \textit{coherent} quantum feedback refers to the situation where the sensors, processors, and actuators are all quantum systems that interact coherently with the quantum system to be controlled \cite{lloydCoherentQuantumFeedback2000a,Jacobs_2014}. In this scenario, the controller receives, processes, and feeds back quantum information. An exciting scientific frontier in this field is the exploration of phenomena that emerge in a regime when the controller can store and process the quantum state of multiple degrees of freedom.

In quantum optical systems, continuous coherent quantum feedback can be introduced naturally by reflecting the output radiation fields of a quantum emitter back onto the emitting system \cite{gardinerInputOutputDamped1985}, e.g., by means of atom-photon interfaces in waveguide QED systems \cite{hoiProbingQuantumVacuum2015}. These kind of coherent feedback loops can acquire a true quantum many-body character when the  associated \textit{time delay} is large, i.e., when the time required for excitations to propagate through the feedback loop is large compared to the time required to emit an excitation and the delay line can accommodate several excitations at a time \cite{pichler_universal_2017}. Remarkably, several recent experiments across multiple platforms can now access this regime of large time delays. For instance, both in optical as well as in microwave settings, new milestones were established in scaling-up distances in distributed quantum networks \cite{Yu2020,Lago-Rivera2021,vanLeent2022,campagne-ibarcqDeterministicRemoteEntanglement2018a,Zhong2019,PhysRevX.5.041044,axlineOndemandQuantumState2018}. Moreover, pioneering experiments with on-chip networks with superconducting devices also accessed this non-Markovian regime by employing slow excitation interconnects realized  with structured waveguides \cite{PhysRevX.11.041043,Chakram2022}, or by using propagating phononic modes \cite{Andersson2019,bienfaitPhononmediatedQuantumState2019,Dumur2021}.  

\begin{figure*}
\includegraphics[width=0.95\textwidth]{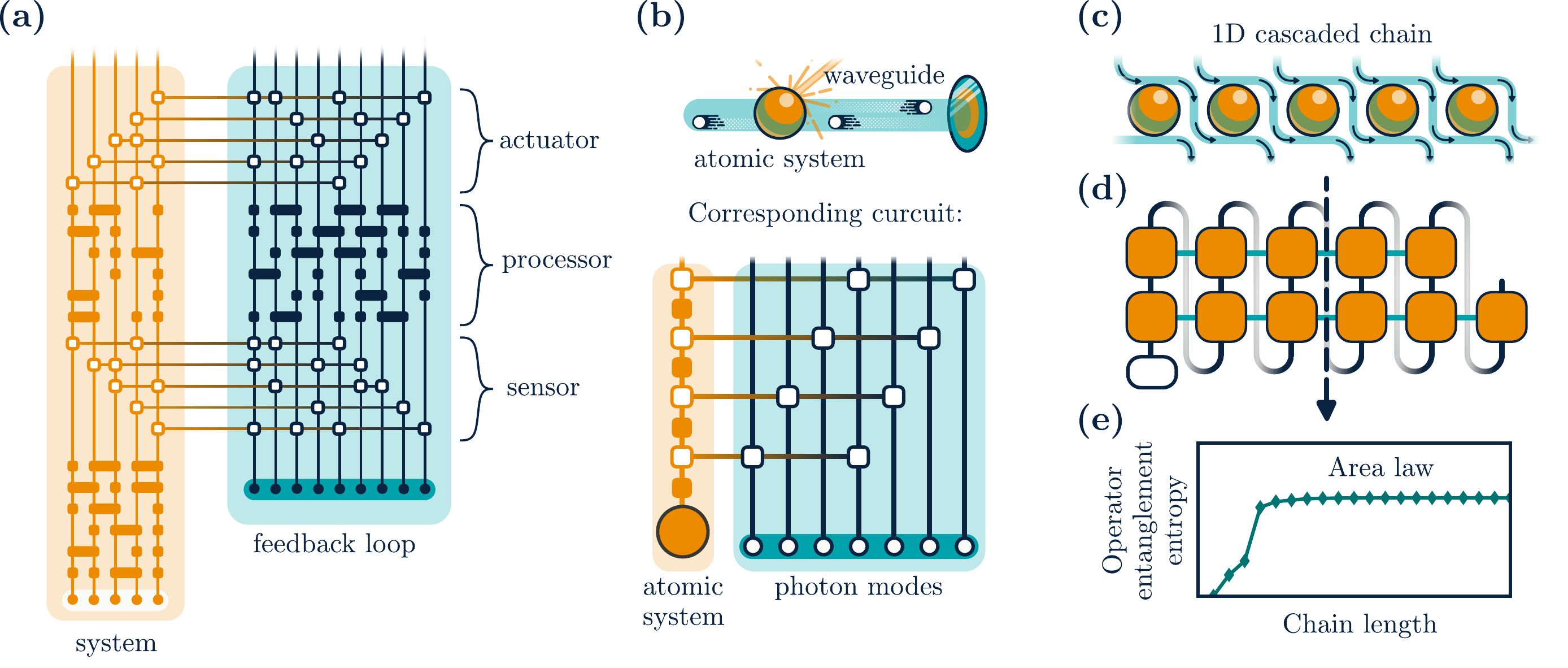}
				\caption{Overview. (a) Schematic depiction of a generic coherent quantum feedback scheme, where a system interacts coherently with a feedback loop consisting of sensor, processor, and actuator. In this general illustration, both systems and feedback loop contain multiple degrees of freedom.  (b) Simplest quantum optical setup that features continuous coherent quantum feedback, where photons propagating from an atomic emitter to a mirror and back represent the feedback loop. If the time delay in this process is large, the feedback loop can host multiple modes. The circuit representation of the dynamics (bottom) highlights the similarities with the generic scheme in (a). Note that in this setting the coherent feedback consists only of a sensor and a (delayed) actuator, since the processor acts trivially.   (c) The problem in (b) can be mapped to the Markovian problem of a 1D cascaded chain, where replicas of the atomic system interact with their nearest neighbors via cascaded channels. (d) A matrix product representation of the evolution operator of the problem. (e) The corresponding operator entanglement entropy obeying an area law.}
    \label{fig:introillustration}
\end{figure*}

On the theoretical side, dealing with time-delayed continuous coherent quantum feedback poses significant challenges, and traditional quantum optical techniques fail: analytical treatments are limited to linear systems \cite{PhysRevA.94.023806,PhysRevA.94.023809,PhysRevA.78.062104} or small excitation numbers in the feedback loops  
\cite{laakso_scattering_2014,PhysRevA.91.053845,PhysRevLett.124.043603,Cai_2021,PhysRevA.106.023708,calajoExcitingBoundState2019,PhysRevA.60.1549,PhysRevA.102.013727}, 
while advanced, non-perturbative techniques \cite{grimsmo_time-delayed_2015,pichler_photonic_2016,guimondDelayedCoherentQuantum2017,PhysRevA.93.062104,whalen_open_2017,PhysRevA.98.012142,PhysRevA.101.023807,PhysRevResearch.3.023168,zhang_embedding_2022,PhysRevE.66.011914,Strathearn2018,PhysRevA.97.012127,PhysRevLett.123.240602,PhysRevX.11.021040,cygorekSimulationOpenQuantum2022,Ye_2021} are typically either limited to finite time delays, or short-time dynamics.
Among the latter, several approaches are based on extending the Markovian cut and including the degrees of the feedback loop using tensor network techniques \cite{pichler_photonic_2016,guimondDelayedCoherentQuantum2017,PhysRevA.93.062104}. However, the computational cost associated with representing the feedback loop typically increases exponentially with the time delay \cite{pichler_photonic_2016}. A similar problem arises in approaches based on tensor network representation of the Feynman-Vernon influence functional \cite{Strathearn2018,PhysRevA.97.012127,PhysRevLett.123.240602,PhysRevX.11.021040,cygorekSimulationOpenQuantum2022,Ye_2021}, which suffers from the growth of temporal entanglement with time delays. An alternative approach is based on representing the system dynamics in the form of a Markovian many-body system \cite{PhysRevE.66.011914,grimsmo_time-delayed_2015,liu2023quantum}. However, applications of this approach suffered from the exponential growth of the many-body Hilbert space, limiting the solution to short-time transient dynamics, preventing the access to steady state quantities \cite{PhysRevA.95.053821}. Predicting properties of systems subject to (continuous) coherent quantum feedback with time delays in generic parameter regimes thus remains an outstanding conceptual challenge. 

In this work we address this challenge and develop methods for efficient and (numerically) exact solutions of the full real time dynamics as well as the steady states values of several important quantities of setups with continuous coherent time delayed quantum feedback.
Below we first illustrate our method in detail on the simplest relevant example, that is, the problem of a single coherently driven two-level system coupled coherently to a long delay line. Based on this example, we review an exact relation between this non-Markovian problem and a corresponding Markovian many-body problem, the one-dimensional (1D) cascaded chain \cite{grimsmo_time-delayed_2015}. This relationship is established in two steps: We first represent the wavefunction of the quantum optical node and of the the delay line as a 2D tensor network \cite{pichler_universal_2017}, and then relate the transfer operator of this tensor network to the propagator of the 1D cascaded chain. 
Our central technical result is that this propagator can be represented accurately in matrix product form, and its operator entanglement entropy obeys an area law in the entire parameter space. We also show that analogous results hold for several additional, more complicated quantum optical problems with time delays.
Leveraging this insight allows us to solve for the real time dynamics and the steady state
 of the reduced state of the system as well as for all low-order correlation functions of the propagating fields. Finally, we develop a semi-analytical approach for the problems studied in this work. This is based on a mean-field approximation of the propagator of the 1D cascaded chain. This approach is motivated by our empirical observation that the effective bond dimension of the propagator is small in large regions of the parameter space. We show that our mean-field ansatz indeed reproduces the exact results in the relevant regions of parameter space.

\section{Model description}\label{sec:model}
\subsection{Continuous coherent quantum feedback with time delays}\label{sec:modelTDF}
\begin{figure}
\subfloat[\label{fig:AtomMirror}]{%
  \includegraphics[width=\columnwidth]{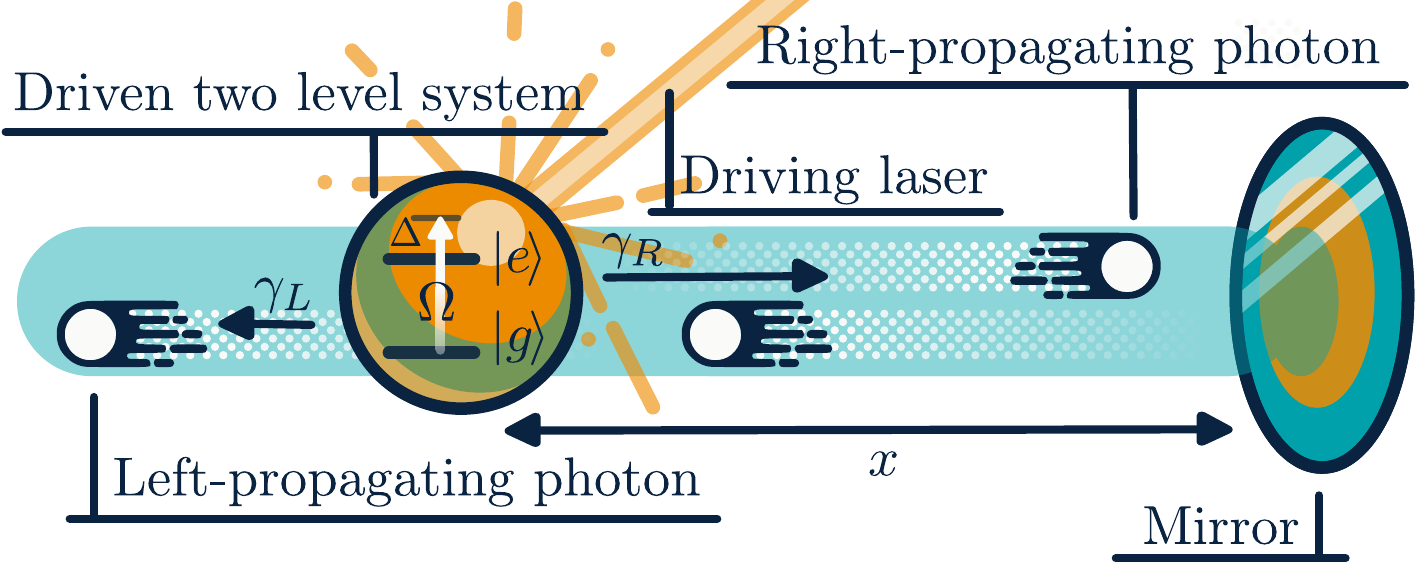}%
}\hfill

\subfloat[\label{fig:Twoatoms}]{%
  \includegraphics[width=\columnwidth]{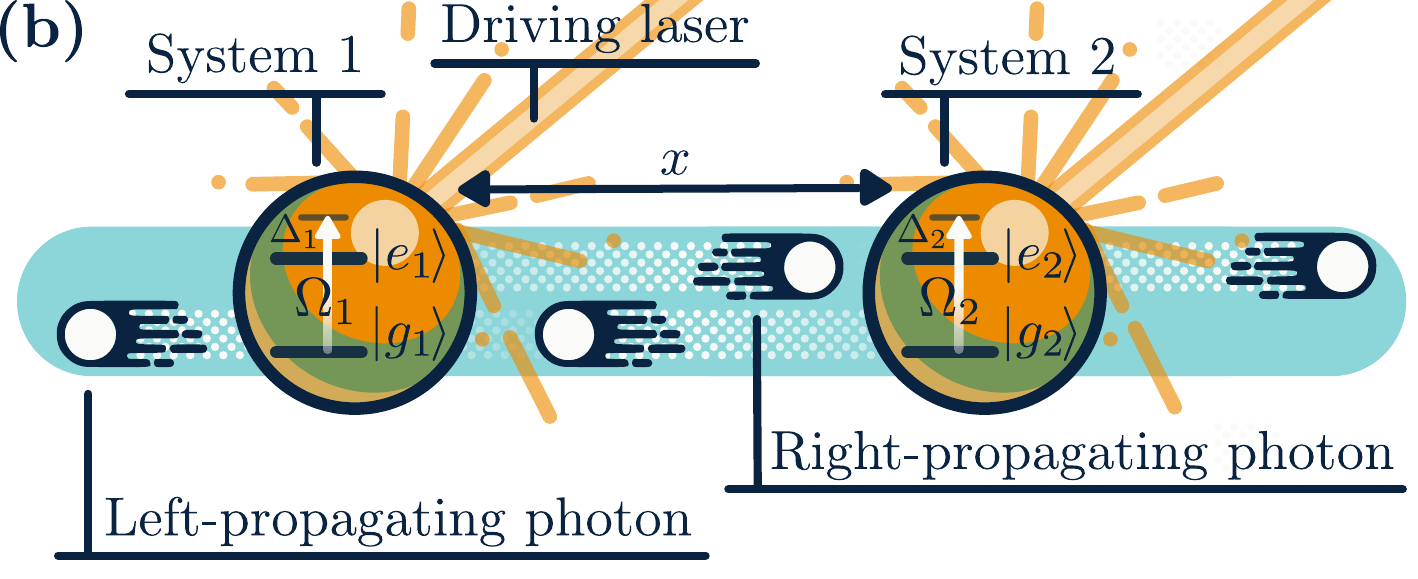}%
}
\caption{Two schematic setups. (a) An atom in front of a mirror: a driven two-level system is coupled to a one-dimensional waveguide, terminated at one side by a mirror. (b) Two driven distant atoms coupled to a one-dimensional waveguide. In both setups, the systems are driven by a classical driving field, either via the waveguide, or via a separate channel (as depicted here).}
\end{figure}

\color{color0}
In this work we develop a new approach to solving problems involving continuous coherent time-delayed quantum feedback. For the sake of clarity, we discuss this approach on 
the simplest but paradigmatic quantum optical model {exhibiting time delays}, consisting of a single driven nonlinear quantum optical system whose output is fed back to itself with a time delay. Physically, this is realized, e.g., by a driven atom coupled to a semi-infinite waveguide with a distant, perfectly reflecting mirror on one side, as shown in Fig.~\ref{fig:AtomMirror}. The total Hamiltonian for this model consists of three terms describing the system (e.g., the atom), the bath (e.g., the waveguide), and their interaction, respectively:
\begin{equation}\label{eqn:totham}
H=H_{\text{sys}}+H_B+H_{\text{int}}.
\end{equation}
For concreteness, below we often use a two-level atom as an example representing the system, where the system Hamiltonian is given by
\begin{equation}\label{eqn:sysham}
H_{\text{sys}}=-\hbar\omega_{eg}\ket{e}\bra{e}-\frac{\hbar}{2}\left(\Omega\ket{g}\bra{e}e^{i\omega_0 t}+\rm h.c.\right).
\end{equation}
Here $\omega_0$ is a driving laser frequency, $\Omega$ is the Rabi frequency, and $\omega_{eg}$ is the atomic transition frequency.  We denote the states of the atom by $\ket{g}$ and $\ket{e}$, and the associated Hilbert space by $\mc{H}_{\rm sys}$. This model can be straightforwardly generalized to higher-dimensional systems, and we denote the system Hilbert space dimension by $d=\dim\left(\mc{H}_{\rm sys}\right)$ in the following.  
The bath Hamiltonian describing, e.g., a 1D semi-infinite waveguide is given by
\begin{equation}
   H_B=\int d\omega\hbar\omega b^{\dagger}(\omega)b(\omega),
\end{equation} 
where 
$b(\omega)\;(b^\dagger(\omega))$ denote a bosonic destruction (creation) operator of a bath excitation  with frequency $\omega$. For convenience we refer to these bath excitations as photons in the following.
To describe the interaction of the system with the one-dimensional waveguide, we define system operators $c_L$ and $c_R$ associated with the coupling to the left- and right-propagating photons and corresponding decay rates $\gamma_L$ and $\gamma_R$. In general, these can be different for left and right moving photons, but for the simple two-level example we chose them to be the same, i.e., we use $c_L=c_R\equiv\ket{g}\bra{e}$ and $\gamma_L=\gamma_R\equiv\Gamma/2$. The Hamiltonian representing the interaction between the system and the bath (in rotating wave approximation) is given by
\begin{multline}
  H_{\text{int}}=\frac{i\hbar}{\sqrt{2\pi}}\int d\omega \left[b^{\dagger}(\omega)\left(c_L\sqrt{\gamma_L}e^{-i\omega x/v}-\right.\right.\\
  \left.\left.-c_R\sqrt{\gamma_R} e^{i\omega x/v}\right)-\rm  h.c.\right],
\end{multline}
where $x$ denotes the distance between the atom and the mirror, and $v$ the photon group velocity in the waveguide with linear dispersion relation. These are connected to the two quantities characterizing the delay line formed by the reflecting waveguide: The delay time  $\tau=2x/v$ required by a photon to propagate from the atom to the mirror and back, and the phase $\phi=\pi-\omega_0\tau$ that a photon with frequency $\omega_0$ accumulates during this round-trip. We note that  couplings of the system to other Markovian environments can be included straightforwardly in this model.

We find it convenient to change from the frequency representation to a time representation of the waveguide radiation modes. For this we introduce the so-called quantum noise operators
\begin{equation}\label{eqn:noiseop}
b(t)=-e^{-i\omega_0\tau/2}\frac{1}{\sqrt{2\pi}}\int d\omega b(\omega)e^{-i(\omega-\omega_0)(t-\tau/2)},
\end{equation}
which satisfy bosonic commutation relations $[b(t),b^\dag (t')]=\delta(t-t')$. The operator $b^\dag (t)$ creates a photon in the radiation mode labelled by $t$. The definition~\eqref{eqn:noiseop} differs from the conventional one \cite{gardinerQuantumWorldUltraCold2015} by a phase and a time shift, which are included here for practical reasons.
With this, the Hamiltonian Eq.~\eqref{eqn:totham} can be rewritten in the frame rotating with the laser frequency $\omega_0$ and in the interaction picture with respect to the bath Hamiltonian as
\begin{equation}\label{eqn:totham2}
H^{R,I}(t)=H^{R,I}_{\text{sys}}+H^{R,I}_{\text{int}}(t).
\end{equation}
\noindent In the example of the driven two-level system, we have $H^{R,I}_{\text{sys}}=-\hbar\Delta\ket{e}\bra{e}-\frac{\hbar}{2}\left(\Omega\ket{g}\bra{e}+\rm h.c.\right)$,
with detuning $\Delta=\omega_0-\omega_{eg}$. The interaction Hamiltonian takes the form
\begin{equation}\label{eqn:intham2}
H^{R,I}_{\text{int}}=i\hbar\left((\sqrt{\gamma_R}b^\dagger(t+\tau)c_R+\sqrt{\gamma_L}b^\dagger (t)e^{i\phi}c_L)-\rm h.c.\right).
\end{equation}
This formulation allows for a transparent interpretation of the dynamics: At each time instant $t$ the system interacts with two modes of the environment, namely the ones labelled by $t+\tau$ and $t$. Note that the modes labelled $s$, with $t<s<t+\tau$, represent the field in the delay line at time $t$, i.e., the radiation field between the atom and the mirror. As time progresses, the system thus interacts with each mode of the environment exactly twice. The time separation $\tau$ between these two events results in a memory of the environment that underlies the non-Markovian nature of this setup. While we derived the model described by eq.~\eqref{eqn:intham2} for the specific setup of an atom coupled to a semi-infinite waveguide, we note that it also applies to other setups, such as giant atoms \cite{friskkockumQuantumOpticsGiant2021} or collisional models \cite{ciccarelloCollisionModelsQuantum2017}. 

\subsection{General quantum optical network}\label{sec:modelgeneral}
This example straightforwardly  generalizes to an arbitrary network of $n$ distant quantum optical nodes interconnected by a set of $w$ photonic channels, which is described by a Hamiltonian of the form 
\begin{align}
H(t)=\sum_{i=1}^n H_{{\rm{sys}}}^{(i)}+ \sum_{i=1}^n\sum_{j=1}^{w_i} H_{{\rm int}}^{(i,j)}(t),
\end{align}
where
\begin{equation}
\!H_{\text{int}}^{x}(t)\!=\!i\hbar\left(\sqrt{\gamma_{x}}b_{\sigma(x)}^\dagger(t+\tau_{x})c_{x}e^{i\phi_{x}}-\rm h.c.\right),
\end{equation} and $x=(i,j)$, a superindex. 
Here a node $i$ is described by a system Hamiltonian $H_{\text{sys}}^{(i)}$ and coupled to $w_i$ photon channels, with jump operators $c_{i,j}$ ($j\in\{1,\dots w_i\}$). The $b_{j}(t)$ are quantum noise operators of the $j$-th waveguides, satisfying $[b_j(t),b_{j'}^\dag(t')]=\delta_{j,j'}\delta(t-t')$. The network structure is completely specified by a set of time delays $\tau_x$, propagation phases $\phi_x$, as well as an index function $\sigma(x)\in \{1,\dots w\}$.
Here and in the following, we drop the superscript $R,I$ (cf. Eqs.~\eqref{eqn:totham2}), since we always work in this frame from now on. 

We note that this model includes the important example of two atoms coupled to a common 1D waveguide at two distant points, see Fig.~\ref{fig:Twoatoms}. In this case, $w=2$, corresponding to the left- ($\sigma=1$) and the right- ($\sigma=2$) moving modes in the waveguide.  Moreover, we stress that the above models also describe so-called giant atoms that can potentially couple at multiple (distant) points to a waveguide \cite{friskkockumQuantumOpticsGiant2021}. Finally, we note that the above model can accommodate standard Markovian channels describing, e.g., the emission of photons into unguided modes. 
Such a general network can be used as a continuous quantum feedback setup, with some of the nodes playing the role of the feedback processor, allowing for more precise control.

\section{Mapping to 1D cascaded chain}\label{sec:mapping1Dchain}

In this section we discuss how the physics of the non-Markovian problem of time-delayed coherent quantum feedback is related to the Markovian problem of the 1D cascaded chain \cite{PhysRevLett.70.2269}. This relationship forms the basis of our numerical algorithm described in Sec.~\ref{sec:nuemricalMethods}. While we illustrate this relationship on the example defined in Sec.~\ref{sec:modelTDF}, the discussion directly generalizes to a subclass of quantum optical networks introduced in Sec.~\ref{sec:modelgeneral}, in particular networks where all time delays are integer multiples of a fundamental time delay $\tau$.

\subsection{Quantum state as 2D tensor network}\label{sec:2DisoTNS}
\subsubsection{Quantum state of system and waveguide}
\begin{figure}
				\includegraphics[width=\linewidth]{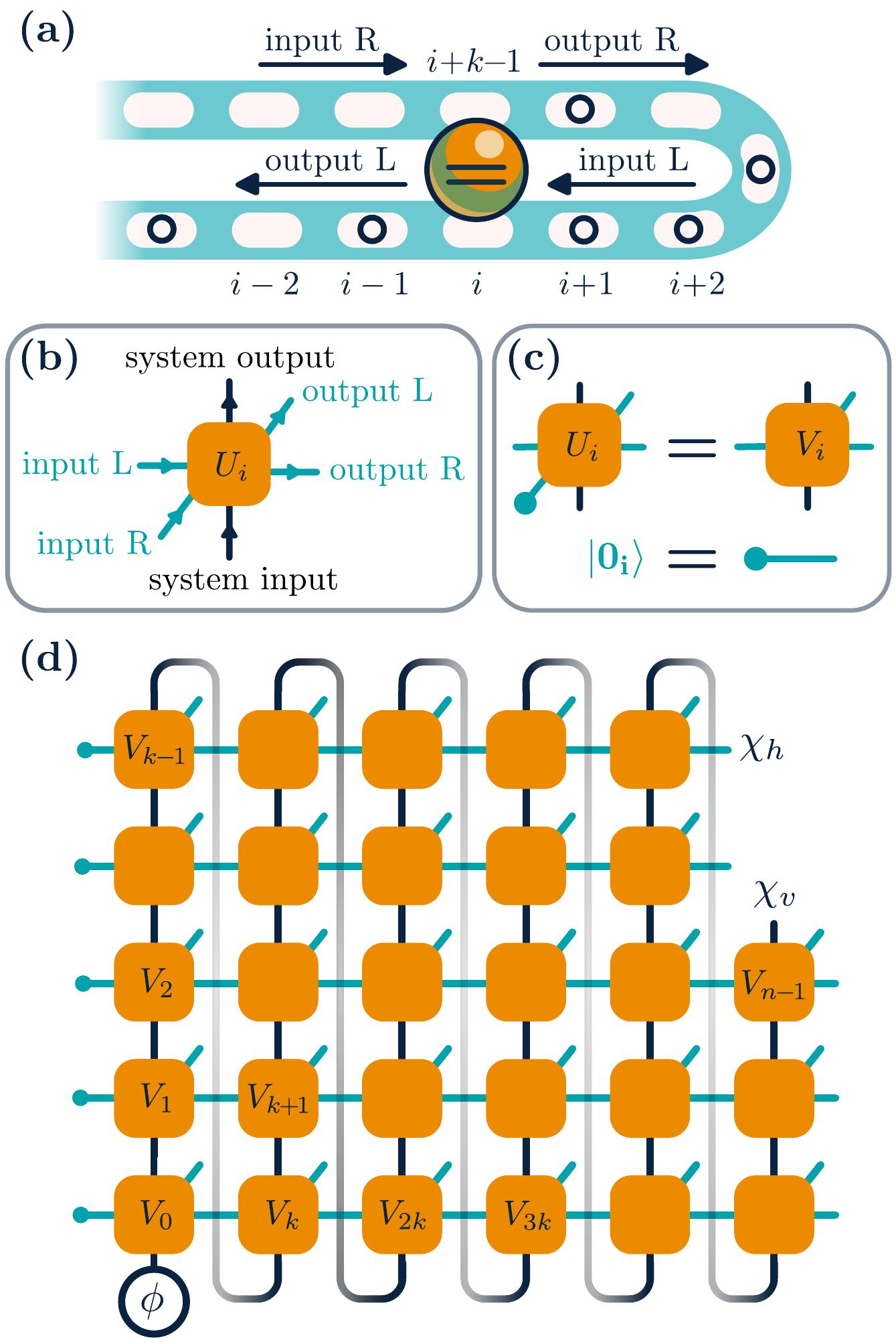}
                \captionsetup{justification=justified}
				\caption{(a) Illustration of the time discretization: The time bins represented by the white bubbles move as a conveyor belt in the directions pointed by the arrows, such that the system simultaneously interacts with the time bins $i$ and $i+k-1$. This interaction imposes a unitary map $U_i$ (b) acting on the atom and the two time bins. It is depicted here as a tensor with three input and three output legs (L and R stand for left- and right-moving  time bins). Right-moving time bins assume to always have a vacuum input, therefore we form an isometry $V_i$ (c). (d) Tensor network representing a total wavefunction of both the system and the waveguide at time $t_n$: The atom is initially in a state $\phi$ and the waveguide including the feedback loop is in a vacuum state. This tensor network corresponds to the equation~\eqref{eqn:isopsi}.
    }
            \label{fig:wavefuntensornetwork}
			\end{figure}

We will now integrate the Schr\"odinger equation associated with the Hamiltonian~\eqref{eqn:totham2} for the total wavefunction of system and waveguide, and introduce a convenient representation of this wavefunction using a tensor network. To start, we formally write the quantum state of the system and bath, i.e., the atom and radiation field, at time $t$ as 
 $
     \ket{\Psi(t)}=U(t-t_0)\ket{\Psi(t_0)},
$
where the evolution operator is
\begin{equation}\label{eqn:evoloperator}
    U(t-t_0)=\mathcal{T}\left\{\exp\left(-\frac{i}{\hbar}\int^t_{t_0}H(t')dt'\right)\right\}
\end{equation}
with $\mathcal{T}$ denoting time-ordering. We consider the initial state of the waveguide with all the modes being in the vacuum state, i.e., $b(t)\ket{\Psi(t_0)}=0$ for all $t$. 
To proceed we find it convenient to discretize the total time of evolution in (infinitesimally) small steps $\Delta t$, such that $t_i=t_0+i\Delta t$ and $\tau=k \Delta t$ with both $i$ and $k$ integer numbers. We define Ito increment operators for each time bin as $
\Delta B_i=\int_{t_i}^{t_i+\Delta t}b(t')dt'.$
 They satisfy the bosonic commutation relations
$[\Delta B^\dagger_i,\Delta B_j]=\Delta t\delta_{ij},$ such that we can associate a (bosonic) Hilbert space $\mathcal{H}_i$ with each time bin $i$ and denote its vacuum state by $\ket{0_i}$. Thus we can Trotterize the evolution operator as $U(t_n-t_0)=U_{n-1}\dots U_{1}U_0$ with 
\begin{equation}\label{eqn:evoloperatortrotstep}
  U_i=\exp\left(-\frac{i}{\hbar}H_{\text{sys}}\Delta t+\Upsilon_i^L+\Upsilon_i^R\right),
\end{equation}
where
\begin{align}\label{eqn:Phi}
\Upsilon_i^R=\sqrt{\gamma_R}c_R\Delta B_{i+k}^\dag -\rm h.c.,\\ \nonumber
\Upsilon_i^L=\sqrt{\gamma_L}e^{i\phi}c_L\Delta B_{i}^\dag  -\rm h.c..
\end{align}
To simplify the following expressions, we introduce the notation $R=\sqrt{\gamma_R}c_R$ and $L=\sqrt{\gamma_L}e^{i\phi}c_L$.
The unitary $U_i$ acts non-trivially only in the Hilbert space of the atom and the Hilbert space of time bins $i$ and 
$i+k$ (cf. Fig.~\ref{fig:wavefuntensornetwork}).
Since the waveguide is initially in the vacuum state, it is useful to form the isometry $V_i: \mc{H}_{\rm sys}\otimes \mc{H}_i\rightarrow \mc{H}_{\rm sys}\otimes \mc{H}_i\otimes \mc{H}_{i+k}$, which is induced by an application of the unitary map to the vacuum state of time bin $i+k$, 
$
  V_i=U_i\ket{0_{i+k}}.
$
With this we can write the state at time $t_n$ as 
\begin{equation}\label{eqn:isopsi}
\ket{\Psi(t_n)}=V_{n-1}\dots V_{1}V_0\ket{\phi}\ket{v}
\end{equation}
where $\ket{\phi}$ is the state of the system at time $t_0$, and $\ket{v}=\bigotimes_{i=0}^{k-1}\ket{0_i}$ is the initial state of the first $k$ time bins, i.e., the initial radiation field in the delay line. For the following discussion we find it useful to depict Eq.~\eqref{eqn:isopsi}  in the form of the tensor network shown in Fig.~\ref{fig:wavefuntensornetwork} (see \cite{pichler_universal_2017} and the Appendix for a detailed discussion). Each isometry $V_i$ corresponds to a tensor in a two-dimensional square lattice. The size of the network along the first  dimension (vertical direction in Fig.~\ref{fig:wavefuntensornetwork}) is set by $k$, i.e., by the round-trip time $\tau$ in units of $\Delta t$, while the size along the second dimension (horizontal direction in Fig.~\ref{fig:wavefuntensornetwork}) is given by $m=\lceil n/k \rceil$, i.e., total evolution time $t_n-t_0$ in units of the  delay time $\tau$ rounded up. The bond dimension along the vertical direction, $\chi_v$, is set by the dimension of the system Hilbert space, $\chi_v = d$, while the bond dimension along the horizontal direction, $\chi_h$, is set by the effective dimension of the bosonic modes associated to each time bin. For our workhorse example of the two-level system in front of the mirror, we have $\chi_v=\chi_h=2$. An important peculiarity of the network geometry are the shifted periodic boundary conditions along the first dimension, as depicted in Fig.~\ref{fig:wavefuntensornetwork}d. We note that this state belongs to the class of 2D isometric tensor network states \cite{wei_sequential_2022,soejima_isometric_2020}, with its orthogonality center located in the lower left corner in Fig.~\ref{fig:wavefuntensornetwork}. The isometric property of the tensors follows here directly from the sequential generation process. 
\begin{figure*}
 \includegraphics[width=0.9\textwidth]{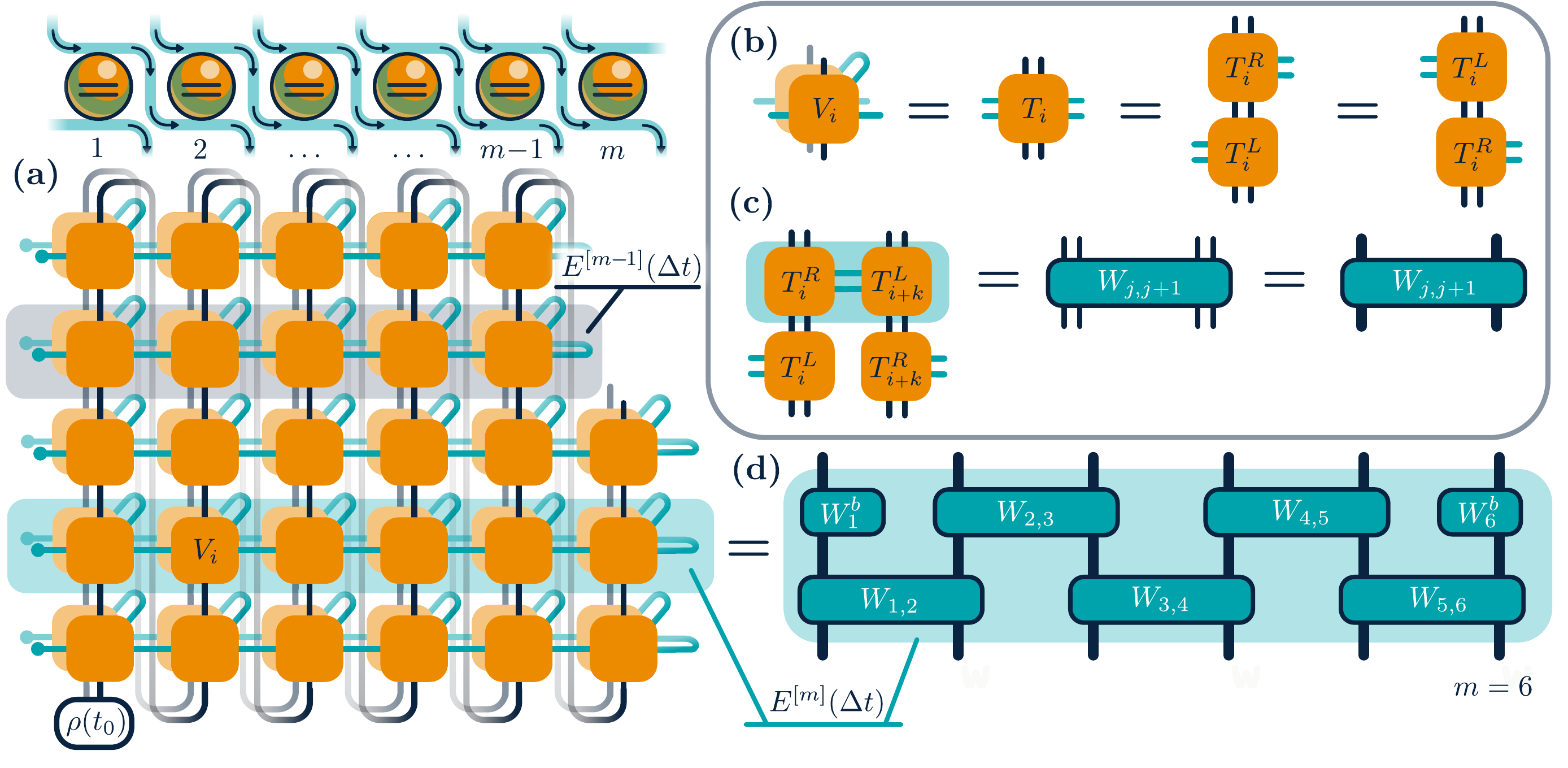}
\caption{(a) To obtain the reduced density operator of the atom, we ``sandwich" the network of the wavefunction $\ket{\Psi(t_n)}$ (Fig.~\ref{fig:wavefuntensornetwork}) with its counterpart  $\bra{\Psi(t_n)}$ and contract all the open legs containing photonic degrees of freedom (light blue legs) to trace out the bath. The transfer superoperator $E^{[m]}(\Delta t)$ for the resulting tensor network is indicated by the light blue area, while the dark blue area encloses $E^{[m-1]}(\Delta t)$. The superoperator $E^{[m]}(\Delta t)$ acts on $m$ replicas shown above the tensor network and consists of $m$ connected tensors $T_i$ (b) acting on the corresponding replica $\lceil i/k \rceil$. These tensors are decomposed as $T_iX=T^L_i T^R_iX$, corresponding to the right output field and the left input field. This decomposition is correct and symmetric up to higher-order Trotter terms: $T_iX=T^R_i T^L_iX$. (c) The tensors acting on the adjacent replicas $T^R_i$ and $T^L_i$, when connected, form a two-site superoperator $W_{j,j+1}=\exp{\mc{L}_{j,j+1}^{\rm casc}\Delta t}$ with $\mc{L}_{j,j+1}^{\rm casc}$ defined by Eq.~\eqref{eqn:Mastereqn}. For simplicity we schematically represent double legs as one thick leg of dimension $d^2$. Tensors on the borders $T^L_1$ and $T^R_{m}$ do not have a connecting pair and result into one-site boundary local propagators $W^b_1=\exp{\mc{L}_1^b\Delta t}$ and $W^b_m=\exp{\mc{L}_m^b\Delta t}$ with the boundary terms defined by Eq.~\eqref{eqn:boundaryterms}. With this we show that the transfer operator $E^{[m]}(\Delta t)$ is an infinitesimal propagator for a 1D cascaded chain (d) as defined in Eq.~\eqref{eqn:deltaE}.} 
    \label{fig:densitymatrixsandwich}
\end{figure*}

\subsubsection{Reduced state of the system}
One of the central quantities of interest is the state of the atom at time $t_n$, described by the reduced density operator $\rho_{\rm sys}(t_n)=\textrm{tr}_{\mc{H}_{B}}\{\ket{\Psi(t_n)}\bra{\Psi(t_n)}\}.$  Here the partial trace is performed over the Hilbert space of the radiation modes, $\mc{H}_B=\bigotimes_{i=0}^{n+k-1} \mc{H}_i$.  For notational simplicity we introduce a superoperator $T_i$, defined via $T_iX=\textrm{tr}_{\mc{H}_i}\{V_i  XV_i^\dagger\}$. This allows us to write 
\begin{align}\label{eqn:rhoTN}\rho_{\rm sys}(t_n)=\textrm{tr}_{\mc{H}_{DL}}\{T_{n-1}\dots T_1T_0 \Phi\},\end{align} where $\mc{H}_{DL}=\bigotimes_{i=n}^{n+k-1} \mc{H}_i$, and $\Phi$ denotes the projector onto the initial state $\ket{\phi}\ket{v}$. In tensor network notation, this expression takes on a simple form shown in Fig.~\ref{fig:densitymatrixsandwich}: It corresponds to a contraction of a network of tensors on a square lattice with shifted periodic boundary conditions. Each tensor in this network corresponds to a map $T_i$, and its dimensions are given by $\chi_h^2$ and $\chi_v^2$ in the horizontal and vertical direction respectively. As illustrated in Fig.~\ref{fig:densitymatrixsandwich} the tensor $T_i$ can be obtained from the tensors $V_i$ and $V_i^\ast$ by contracting the leg corresponding to the output field of the $i$th time bin.
\subsection{Relation to cascaded chain}\label{sec:1Dcascchain}
To gain insight into the 2D tensor network defined by Eq.~\eqref{eqn:rhoTN}, it is useful to consider its transfer operator $E\strut^{[m]}\!(\Delta t)$, as defined in Fig.~\ref{fig:densitymatrixsandwich}.  This transfer operator is a map from the $m$-fold replicated Hilbert space of system operators onto itself, i.e., $E\strut^{[m]}\!(\Delta t): \mc{B}(\mc{H}_{\rm sys})^{\otimes m}\rightarrow \mc{B}(\mc{H}_{\rm sys})^{\otimes m}$. Importantly, it can be shown that $E\strut^{[m]}\!(\Delta t)$ can be exactly written as the (infinitesimal) propagator generated by a Lindblad superoperator, $\mc{L}\strut^{[m]}$, acting on these $m$ replica systems, that is
\begin{align}\label{eqn:deltaE}
    E\strut^{[m]}\!(\Delta t)=\exp\left(\Delta t\mc{L}\strut^{[m]}\right).
\end{align}
Specifically, $\mc{L}\strut^{[m]}$ is the Lindblad superoperator describing the dynamics of the \textit{1D cascaded chain} of the $m$ replica systems described by the system Hamiltonian, such as in Eq.~\eqref{eqn:sysham}, i.e., 
\begin{align}\label{eqn:bigLinbl}
\mc{L}\strut^{[m]}=\mc{L}_1^{b}+\sum_{j=1}^{m-1}\mc{L}^{\rm casc}_{j,j+1}+\mc{L}_m^{b}.
\end{align}
Here $\mc{L}^{\rm casc}_{\scriptscriptstyle j,j+1}$ is the Lindblad operator corresponding to a \textit{cascaded coupling} between replicas $j$ and $j+1$. Such cascaded coupling has been studied first by Gardinder and Carmichael \cite{PhysRevLett.70.2269,carmichaelQuantumTrajectoryTheory1993}, and more recently in the context of chiral quantum optical systems \cite{PhysRevA.105.023712}. Cascaded couplings arise when an output field of a system (e.g., replica $j$) is injected as input to another system (e.g., replica $j+1$) via a unidirectional channel. In the 1D cascaded chain, nearest neighbors are coupled in this unidirectional manner (see Fig.~\ref{fig:densitymatrixsandwich} for an illustration). Mathematically, the Lindblad operator describing the cascaded interaction between replicas $j$ and $j+1$ is given by (cf.
\cite{gardinerQuantumWorldUltraCold2015})
\begin{align}\label{eqn:Mastereqn}
\mathcal{L}^{\rm casc}_{j,j+1}X\!&=\!-\frac{i}{\hbar}\left[H_{j,j+1}^{\rm casc},X\right]+\mathcal{D}[R_j+L_{j+1}]X,
\end{align}
where we defined the cascaded Hamiltonian 
\begin{align*}
    H_{j,j+1}^{\rm casc}=\frac{1}{2}\lr{H_{\text{sys},j}+H_{\text{sys},j+1}+i\lr{ R_j^\dag L_{j+1}-L_{j+1}^\dag R_j}}
\end{align*}
and introduced the shorthand notation $\mc{D}[C]X=CXC^\dag-\frac{1}{2}(C^\dag C X+XC^\dag C)$. Here $H_{\text{sys},j}$, $L_j$ and $R_j$ are simply the system Hamiltonian (e.g., Eq.~\eqref{eqn:sysham}) and jump operators, acting on the $j$th replica system. 
Note that the total Lindblad operator \eqref{eqn:bigLinbl} also contains the boundary terms which are simply given by 
\begin{equation}\label{eqn:boundaryterms}
\begin{aligned}
\mc{L}_1^{\textrm{bd} L}X&=-\frac{i}{2\hbar}\left[H_{\text{sys},1},X\right]+\mathcal{D}[L_1]X,\\
\mc{L}_m^{\textrm{bd} R}X&=-\frac{i}{2\hbar}\left[H_{\text{sys},m},X\right]+\mathcal{D}[R_m]X,
\end{aligned}   
\end{equation}
and act independently only on the first and the last replica. We refer the reader to Fig.~\ref{fig:densitymatrixsandwich} for a diagramatic derivation of this equivalence between the transfer operator $E\strut^{[m]}\!(\Delta t)$ and the propagator of the 1D cascaded chain. A formal derivation can be found in Appendix \ref{app:1Dchain}. This correspondence has an intuitive physical origin already pointed out in Ref.~\cite{grimsmo_time-delayed_2015}: The right-propagating output field emitted by the system at a time $s$ becomes the left-propagating input field of the system at a later time $s+\tau$. In turn, the right-propagating output field of the system at time $s+\tau$ turns into the left-propagating input field of the system at time $s+2\tau$ etc. The different replicas in the cascaded chain thus assume a role analogous to the one of the system at different points in time, separated by multiples of $\tau$.
From this equivalence between the tensor network transfer operator and the infinitesimal propagator of the 1D cascaded chain, it is straightforward to see that the reduced state of the system, $\rho_{\rm sys}(t_n)$, can be obtained from the finite time propagators $E\strut^{[m]}(s)=\exp\left(s\mc{L}\strut^{[m]}\right)$. To be specific, we define $r$ via $t_n=(m-1)\tau+r$, with $0\leq r\leq\tau$. As shown in Fig.~\ref{fig:densitymatrixsandwich}a, the reduced state of the system, $\rho_{\rm sys}(t_n)$, can be obtained from contracting $E\strut^{[m]}\!(r)$ with $E\strut^{[m-1]}\!(\tau-r)$, with shifted periodic boundary conditions, and applying the resulting composite map to the initial state of the system, $\rho_{\rm sys}(t_0)$. Denoting the contraction imposed by shifted periodic boundary conditions applied to a tensor network $X$ by $\mc{P}(X)$, we can write \begin{align}\label{eq:SPC}
    \rho_{\rm sys}(t_n)=\mc{P}(E\strut^{[m-1]}\!(\tau-r)E\strut^{[m]}\!(r))\rho_{\rm sys}(t_0).
    \end{align}
This contraction can be conveniently performed if $E\strut^{[m]}\!(s)$ is given in matrix product form, as discussed in the next section.
\begin{figure*}
\includegraphics[width=0.9\textwidth]{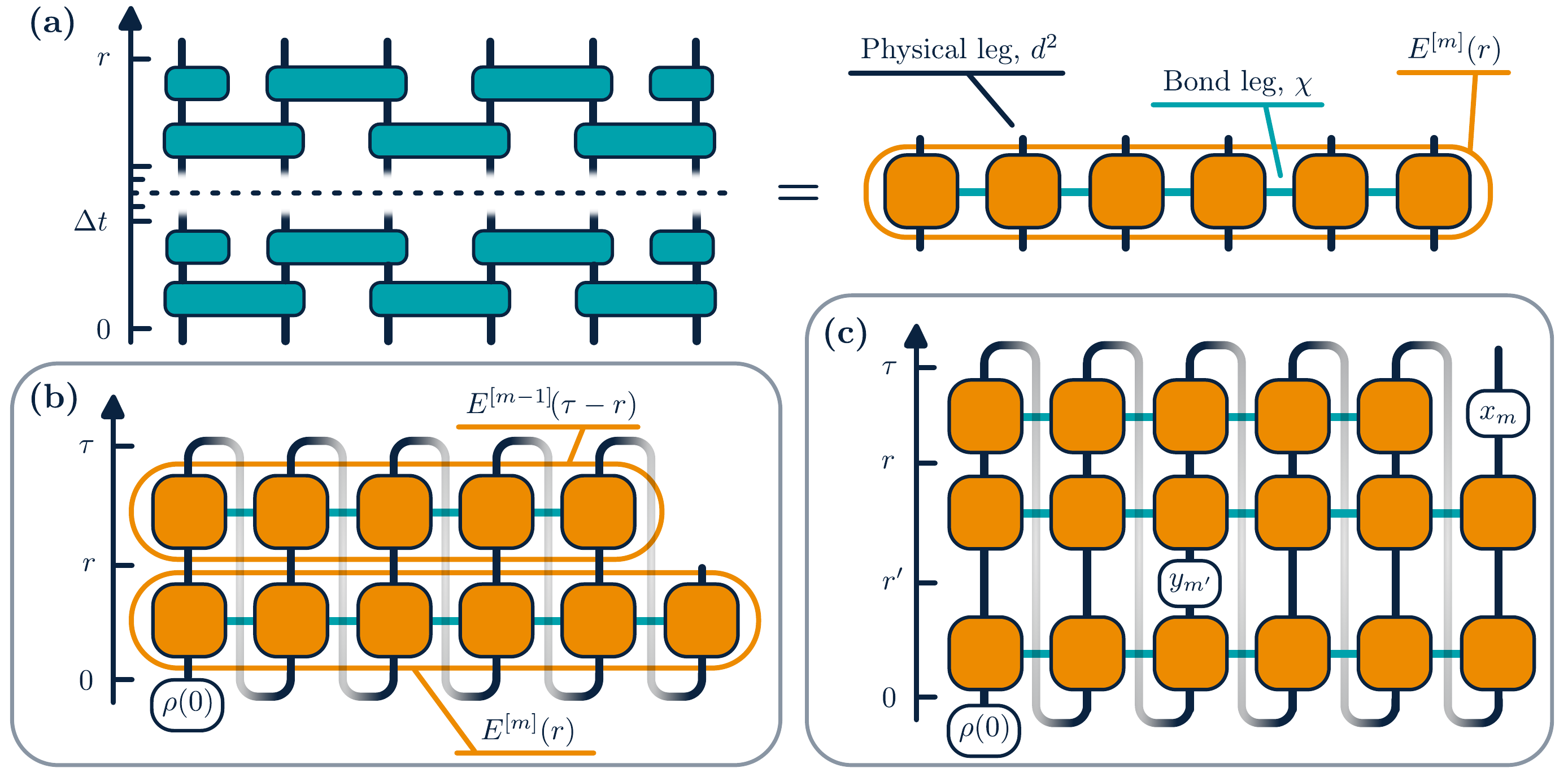}
				\caption{Tensor network illustration of the numerical method described in Sec.~\ref{sec:nuemricalMethods}: (a) The propagator $E\strut^{[m]}(r)$ is obtained by the Trotterized evolution of $m$ sites until time $r$. (b) The reduced density matrix of the atom $\rho_{\rm sys}(t_n)$ is a result of contraction of the propagators  $E\strut^{[m]}(r)$ and $E\strut^{[m-1]}(\tau-r)$. (c) The two-times correlation function of the system operators $\left<x (t)y(t')\right>$ requires a contraction of the three propagators with the system operators inserted between them at the corresponding times and sites.}
    \label{fig:contraction}
\end{figure*}

\subsection{Multi-node networks}\label{Sec:MappingMulti}
The mapping of the delayed quantum feedback problem to the 1D cascaded chain described above can be generalized to more complicated  networks with multiple nodes. In this work we restrict ourselves to networks where all time delays between nodes are identical. In this case one can show that the reduced state of the nodes is always described by a 2D tensor network (with generalized shifted periodic boundary conditions), whose transfer operator is the propagator of a 1D unidirectional master equation on several replicas of the nodes. An important example in this class is the problem of two distant nodes, $A$ and $B$, interacting with a common, bidirectional waveguide, where the time delay due to the photon propagation between the two nodes in both directions is identical: For large time delays the node dynamics is 
non-Markovian, but again it maps to the Markovian problem of a 1D cascaded chain with a two-site unit cell where every odd site corresponds to a replica of node $A$ and every even site to a replica of node $B$.
We refer the reader to Appendix~\ref{app:multiatom} for a detailed derivation of this correspondence. 
\section{Numerical methods}\label{sec:nuemricalMethods}

\subsection{Matrix product form of the propagator}
We are aiming now to efficiently represent the propagator of the 1D cascaded chain, $E\strut^{[m]}(s)=\exp\left(s\mc{L}\strut^{[m]}\right)$  for $0\leq s\leq \tau$, using matrix product state (MPS) techniques. 
For this we first recall that $E\strut^{[m]}(s)$ is a linear operator which maps the Liouville space of $m$-fold replicated system operators, $\mc{B}(\mc{H}_{\rm sys})^{\otimes m}$, onto itself. Following the literature (e.g., Ref.~\cite{gardinerQuantumWorldUltraCold2015}), we refer to such maps as superoperators. Note that such superoperators form a $d^{4m}$-dimensional vector space. Defining  $D=d^4$, this vector space is the tensor product space of $m$ $D$-dimensional vector spaces, each a local superoperator vector space $\mc{C}$, whose elements act only on one site of the 1D chain and map $\mc{B}(\mc{H}_{\rm sys})$ onto itself. The tensor product space is therefore $\mc{C}_m=\mc{C}^{\otimes m}$.
We choose a basis of $\mc{C}$ and denote its basis elements by $S_i$, with $i=1,\dots, D$, from which we can construct a product basis of $\mc{C}_m$. 
With this we can write any superoperator $\mc{S}\in \mc{C}_m$ in a matrix product form
\begin{equation}\label{eqn:MPSO}
\mc{S}=\!\!\!\!\!\!\sum_{j_1,j_2,\dots, j_m=1}^{D}\!\!\!\!\!\! C^{[1]}_{j_1}C^{[2]}_{j_2}\cdots C^{[m]}_{j_m}S_{j_1}\otimes S_{j_2}\otimes\cdots \otimes S_{j_m},
\end{equation}
where $S_{j_i}$ is a local basis superoperator on the site $i$ and the summation includes all such superoperators. The matrix $C^{[i]}_{j_i}$ associated with the local basis superoperator $S_{j_i}$ has dimension $\chi\times\chi$, with $\chi\geq 1$ being the bond dimension of $\mc{S}$ (the boundary tensors $C^{[1]}_{j_1}$ and $C^{[m]}_{j_m}$ are simply vectors of length $\chi$).
We refer to a superoperator in the above form as matrix product superoperator (MPSO).
\subsubsection{Evolution equation}
To construct a representation of the finite time propagator $E\strut^{[m]}(s)$ in the matrix product form Eq.~\eqref{eqn:MPSO}, we first recall that it satisfies
\begin{align}\label{eq:Prop_differential}
\frac{d}{ds}E\strut^{[m]}(s)=\mc{L}^{[m]}E\strut^{[m]}\!(s),
\end{align}
with the initial condition $E\strut^{[m]}(0)=\mathbb{1}^{\otimes m}$. Importantly, $E\strut^{[m]}(0)$ is a product (super)operator. Moreover, $\mc{L}\strut^{[m]}$ contains only nearest-neighbor terms. Therefore, we can use the  standard  time-evolving block decimation (TEBD) algorithm \cite{schollwoeck_density-matrix_2011} to integrate Eq.~\eqref{eq:Prop_differential}. For this we Trotterize the propagation with the cascaded Lindbladian for an infinitesimal time step into $m-1$ nearest-neighbor propagators, $W_{j,j+1}=\exp{\left(\Delta t \mc{L}_{j,j+1}^{\rm casc}\right)}$, and two local boundary terms $W_1^b=\exp(\Delta t \mc{L}_1^b)$ and $W_m^b=\exp(\Delta t \mc{L}_m^b)$ (see Fig.~\ref{fig:densitymatrixsandwich}(d) and Fig.~\ref{fig:contraction}(a)). The computational cost of the associated updates in the matrix product representation for each such two-site update is $O(\chi^3d^{12})$. Note that the time step $\Delta t$ in this Trotterization has to be chosen much smaller than the timescale on which the system evolves, e.g., $\Delta t\ll 1/|\Omega|, 1/|\Delta|, 1/\Gamma$. In all our numerical results below, we checked convergence in the size of $\Delta t$. 
\subsubsection{System density operator}
Once the propagators $E\strut^{[m]}(s)$ are obtained in matrix product form, the reduced state of the system can be calculated at all times $t\leq m\tau$ via Eq.~\eqref{eq:SPC}. For this, first note that one can obtain $E\strut^{[m-1]}(s)$ directly from $E(m,s)$: Due to the unidirectional nature of a cascaded chain one can simply trace out the $m$th replica to obtain the propagator for a shorter chain, that is $E\strut^{[m-1]}(t)=\frac{1}{d}\ptr{m}{E\strut^{[m]}(t)}$. The contraction \eqref{eq:SPC} can then be performed efficiently, since it can be cast in the form of a 1D tensor network contraction as shown in Fig.~\ref{fig:contraction}(b). The computational cost of this contraction is  $O(m\chi^3 d^4)$. 
\subsubsection{Multi-time correlation functions}
Beside the system density operator, we are also interested in the properties of the radiation field. Importantly, arbitrary field correlation functions can be related to multi-time correlation functions of system operators, using input-output relations  (see Ref.~\cite{gardinerQuantumWorldUltraCold2015}). Multi-time correlation functions can be accessed by a straightforward generalization of the above discussion.
For instance, consider the two-times correlation function $\left<x(t)y(t')\right>$ for two arbitrary system operators $x$ and $y$. We introduced the notation $\mean{\dots}=\bra{\Psi(t_0)}\dots\ket{\Psi(t_0)}$ for quantum mechanical expectation values and denote the system operator $x$ in the Heisenberg picture at time $t$ by   $x(t)=U^\dag(t-t_0)xU(t-t_0)$. Without loss of generality we consider $t>t'$. We define integers $m$ and $m'$ as well as remainders  $r$ and $r'$, via $t=(m-1)\tau+r$ and  $t'=(m'-1)\tau+r'$.  We then can write $\left<x (t)y(t')\right>=\tr{\mc{P}(M)\rho(0)}$ with 
\begin{align}\label{eq:mt1}
    M=E\strut^{[m-1]}(\tau-r)x_{m}E\strut^{[m]}(r-r')y_{m'}E\strut^{[m]}(r')
\end{align}
for $r\geq r'$, and 
\begin{align}\label{eq:mt2}
    M=E\strut^{[m-1]}(\tau-r')y_{m'}E\strut^{[m]}(r'-r)x_{m}E\strut^{[m]}(r)
\end{align}
for $r<r'$. Here $x_m$ denotes the operator $x$, acting on the $m$th replica, and $y_{m'}$ is defined analogously (see Fig.~\ref{fig:contraction}c).
Again this contraction can be performed efficiently with a cost given by $O(m\chi^4 d^4)$. 
This can be straightforwardly generalized to arbitrary $p$-times correlation functions of system operators. It is easy to see that the corresponding contraction can be performed at a cost $O(m\chi^{p+2} d^4)$, leading to an exponential scaling with the order of the correlation function $p$. 
\begin{figure*}
\includegraphics[width=0.85\textwidth]{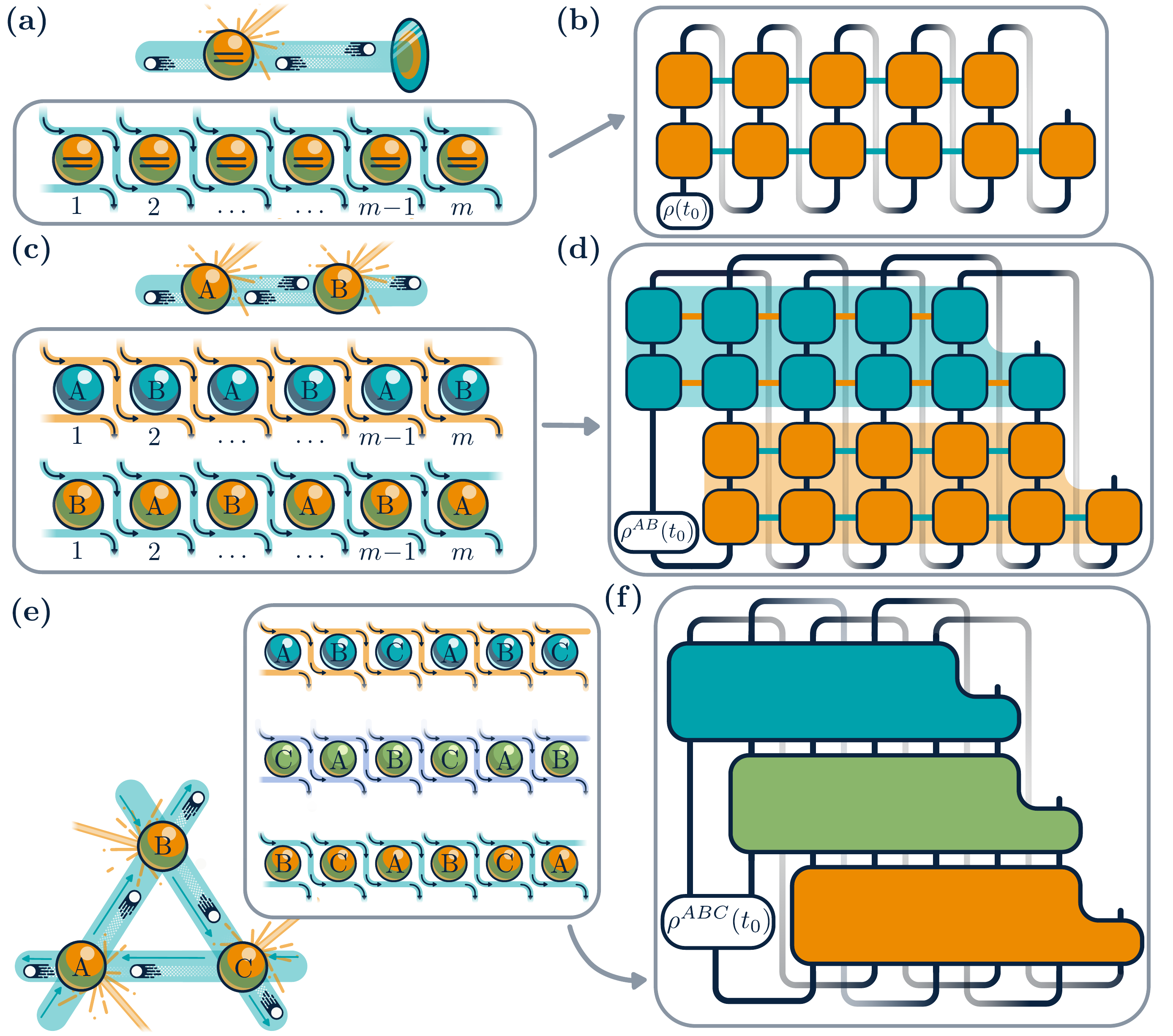}
				\caption{Generalization of our approach to the multinode networks: (a) The atom in front of the mirror is mapped to 1D cascaded chain of identical replicas corresponding to the state of the atom at a different time. The chain is evolved using MPS methods to obtain the total propagator, which is then contracted (b) with the shifted periodic boundary conditions and applied to the initial density operator. (c) A problem of the setup with two atoms connected through a bidirectional waveguide is mapped to the two cascaded chains. (d) The propagators obtained after the evolution of these chains are then contracted together with the double-shifted periodic boundary conditions. (e) A configuration with three atoms connected through a unidirectional waveguide is mapped to the three cascaded chains. (f) The resulting three propagators are contracted  with the triple-shifted periodic boundary conditions.}
    \label{fig:multinodes}
			\end{figure*}
 
\subsection{Infinite cascaded chain}\label{sec:infchain}
One of the most important quantities of interest is the steady state of the system density operator, $\rho_{\rm ss}=\lim_{t\rightarrow \infty} \rho(t)=\lim_{m\rightarrow \infty} \rho(m\tau)$. If the steady state is unique, it can be expressed in terms of the finite-time propagator of an infinite 1D cascaded chain (see Eq.~\eqref{eq:SPC}) as \begin{align}\label{eqn:rhoss}\rho_{\rm ss}=\lim_{m\rightarrow \infty}\mc{P}(E\strut^{[m]}(\tau))\frac{\mathbb{1}}{d}.\end{align} Importantly, we can directly target the steady state of the system by directly calculating infinite system size propagator using infinite matrix product state techniques \cite{PhysRevLett.98.070201}. For this, one assumes a translation invariant 
matrix product representation of $E\strut^{[\infty]}(s)$, with $C^{[k]}(s)=C(s)$ (for all $k$), and integrates Eq.~\eqref{eq:Prop_differential} self-consistently from $s=0$ to $s=\tau$ \footnote{In practice, following standard iTEBD algorithms, this is implemented using a two-site ansatz,  $C^{[2k]}(s)=A(s)$ and $C^{[2k+1]}(s)=B(s)$ (for all $k$).}.  Besides accessing  $\rho_{\rm ss}$, this enables the calculation of multi-time system correlation functions in the steady state using expressions analogous to Eqs.~\eqref{eq:mt1} and \eqref{eq:mt2}, and, in turn, field correlation functions via input-output relations. Moreover, one can also directly access the relaxation time $t_{\rm ss}$ of the system: This is determined by the correlation length $\xi$ of $E^{[\infty]}(\tau)$, via $t_{\rm ss}=\xi\tau$. Since $E^{[\infty]}$ is given in translational invariant matrix product form, its correlation length can be directly accessed from spectral decomposition of the tensor $C^{[k]}(\tau)$. For details regarding the infinite chain algorithm, we refer the reader to Appendix~\ref{app:infchain}.

\subsection{Multi-node networks}
As discussed in Sec.~\ref{Sec:MappingMulti}, we consider $n$-node networks where all time delays are identical. These can also be mapped to 1D cascaded chains. Therefore, we can straightforwardly generalize  the numerical methods introduced above to such multi-network setups. Fig.~\ref{fig:multinodes} illustrates this generalization for  setups with two and three atoms. In each case the tensor network representing the reduced state of the $n$ nodes can be constructed from the propagator of a 1D cascaded chain with an $n$-site unit cell. Specifically, as shown in Fig.~\ref{fig:multinodes}(b) and (c), the state of the nodes is obtained from a contraction of $n$ such propagators with $n$-fold shifted periodic boundary conditions. For each 1D cascaded chain we construct this propagator using standard TEBD procedure in the exact same way as for the case of a 
 single node. The computational cost of constructing the propagator is independent of the number of nodes $n$. However, the cost of contraction scales as $O(m \chi^{3+n}d^{4+2n})$, where $n$ is the number of the nodes in the model, and we assume that all nodes have the same local  Hilbert space dimension $d$. 
\section{Results}\label{sec:results}
\subsection{Propagator bond dimension}

   \begin{figure}
   \subfloat[\label{fig:entropyarealaw}]{%
\includegraphics[width=0.5\columnwidth]{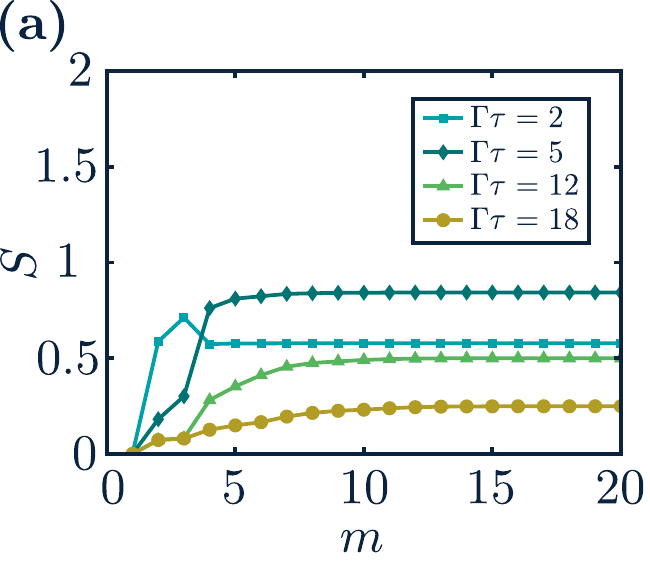}%
}\hfill
\subfloat[\label{fig:entropyphi}]{%
  \includegraphics[width=0.5\columnwidth]{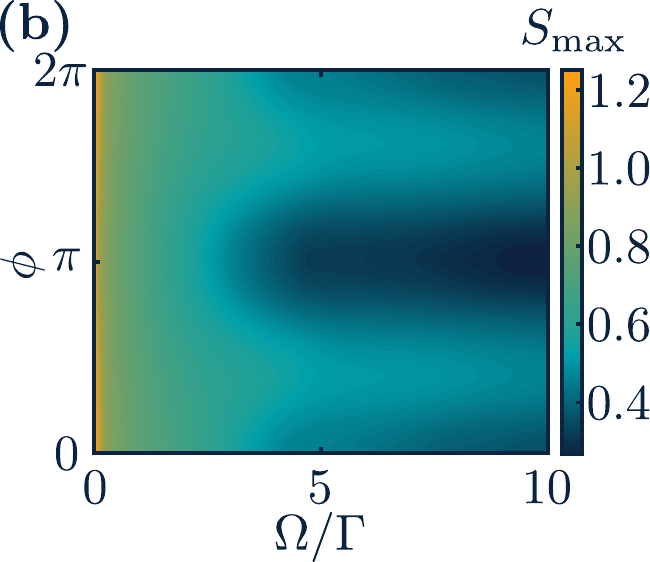}%
}\vfill
   \subfloat[\label{fig:schmidtval}]{%
\includegraphics[width=0.5\columnwidth]{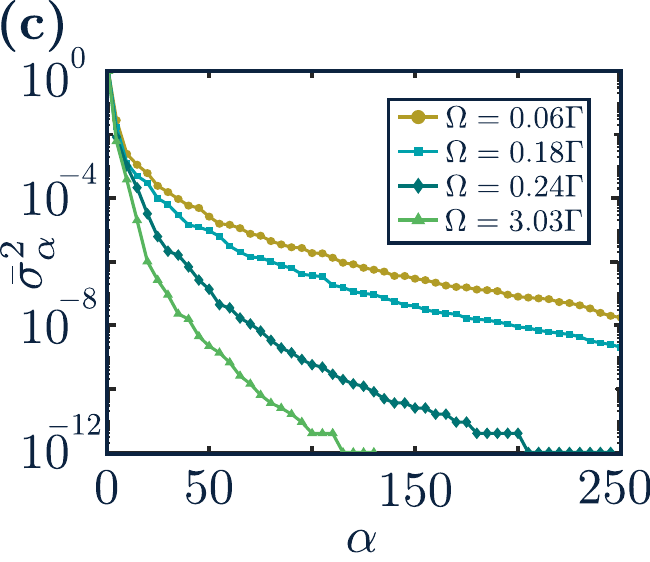}%
}\hfill
\subfloat[\label{fig:entropy0}]{%
  \includegraphics[width=0.5\columnwidth]{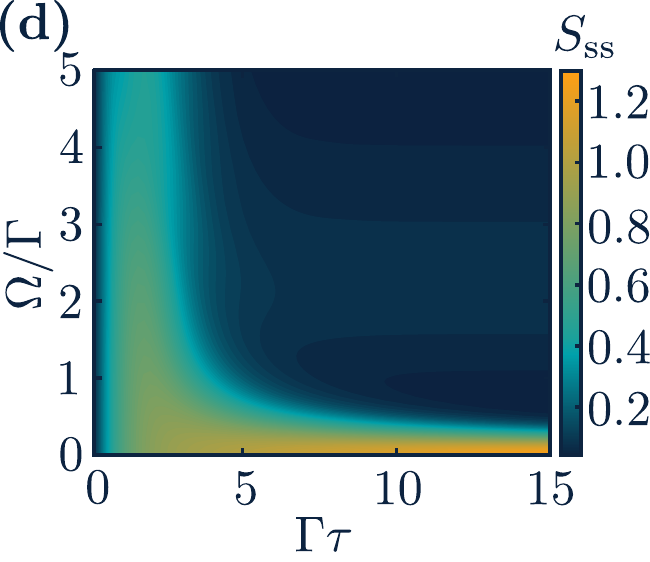}%
}
\caption{Entropy area law demonstrated for the propagator of the atom in front of the mirror: (a) Dependence of the bipartite entropy $S$ on the length of the chain $m$ exhibiting an area law for $\Gamma\tau=2,5,12,18$ and $\Omega/\Gamma=0.5$, $\chi=50$ and Trotter step $\Gamma\Delta t=0.05$. (b) The dependence of the maximum entropy during the time evolution, $S_{\rm max}=\max_\tau S$, on the phase and the Rabi frequency for $m=20$, $\chi=50$ and $\Gamma\Delta t=0.05$. (c) Steady state normalized singular values squared corresponding to the maximum entropy bipartition for different driving strengths, $\Gamma\tau=20$, $\phi=0$, $\chi=250$, $\Gamma\Delta t=0.1$. (d) Steady state entropy $S_{\rm ss}$ for different delay times $\Gamma\tau$ and Rabi frequencies, $\phi=\pi$, $\chi=100$, and  $\Gamma\Delta t=0.001$. For all plots the detuning $\Delta/\Gamma=0$.}
\label{fig:entropy}
\end{figure}
We now proceed to discuss the computational cost of the method  outlined in the previous section, i.e., the cost of constructing the propagator of the 1D cascaded chain. The computational cost depends crucially on the bond dimension of $E\strut^{[m]}(\tau)$: The problem of interest can be solved efficiently if the matrix product representation of $E\strut^{[m]}(\tau)$ obeys an area law for all $\tau$, that is, if the bond dimension $\chi$ required to represent the $E\strut^{[m]}(\tau)$ grows at most polynomially with $m$. Important quantities in this context are the singular values of the splitting of $E\strut^{[m]}(\tau)$ in two partitions formed by the  first $\ell$ replicas and the last $m-\ell$ replicas, respectively. We denote these singular values by $\sigma_\alpha$ (with $\alpha=1,\dots,\chi$), and define  normalized singular values as $\bar\sigma_\alpha=\sigma_\alpha/(\sum_{\beta}\sigma_\beta^2)^{1/2}$. We also introduce the entropy of the normalized singular values associated with this splitting 
\begin{equation}\label{eqn:entropy}
S(\ell)=-\sum_\alpha\bar{\sigma_\alpha}^2\log_2(\bar{\sigma_\alpha}^2),
\end{equation}
as well as the maximum entropy among all cuts of the chain $S=\max_{\ell}S(\ell)$, and use it as a proxy for the bipartite correlations in the propagator and the effective bond dimension $\chi\sim 2^S$. 

In Fig.~\ref{fig:entropy}(a),  we show $S$ as a function of $m$ for the 1D cascaded chain corresponding to our example of a driven two-level atom coupled to a delay line (see Sec.~\ref{sec:modelTDF}). Importantly, this shows a clear \textit{area law} for all values of $\tau$, as the entropy saturates to a finite value as $m$ increases. In Fig.~\ref{fig:entropy}(b) we show this saturation value, calculated for an infinite number of replicas, and confirm that the saturation value is finite in the entire  parameter space, demonstrating the applicability of our method even in previously inaccessible regimes. Perhaps counterintuitively, the largest  entropies are observed for weak driving, while the entropy is remarkably small if both $\tau$ and $\Omega$ are large. We will use this feature in Sec.~\ref{sect:meanfield} and propose a semi-analytical approach to describe the system in this latter regime.

While our results in Fig.~\ref{fig:entropy} demonstrate the area law explicitly for the 1D cascaded chain of driven two-level systems, we found analogous results also for other examples. In general, one expects an area law for the propagator of a 1D Markovian master equation whenever it is rapidly mixing, i.e., its mixing time scales at most logarithmic with $m$.  \cite{Brandao2015}.

\subsection{Single driven atom in front of a distant mirror}

 \begin{figure}[b]
   \subfloat[\label{fig:1atom}]{%
  \includegraphics[width=0.5\columnwidth]{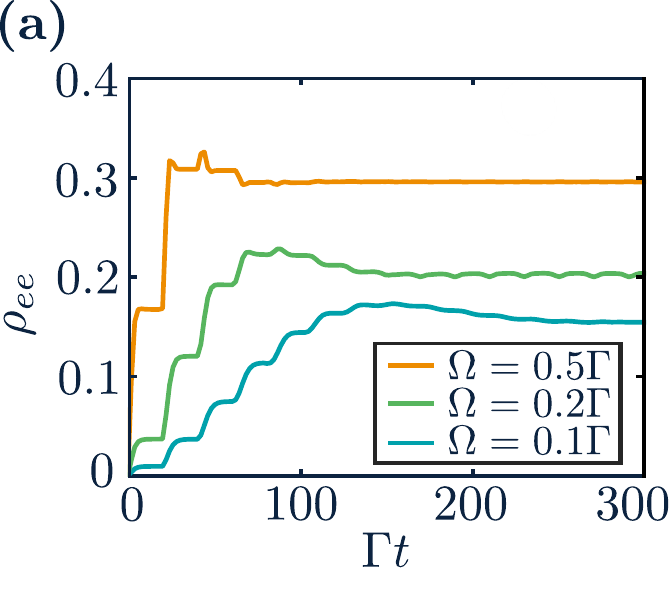}%
  }\hfill
\subfloat[\label{fig:rhoee}]{%
  \includegraphics[width=0.5\columnwidth]{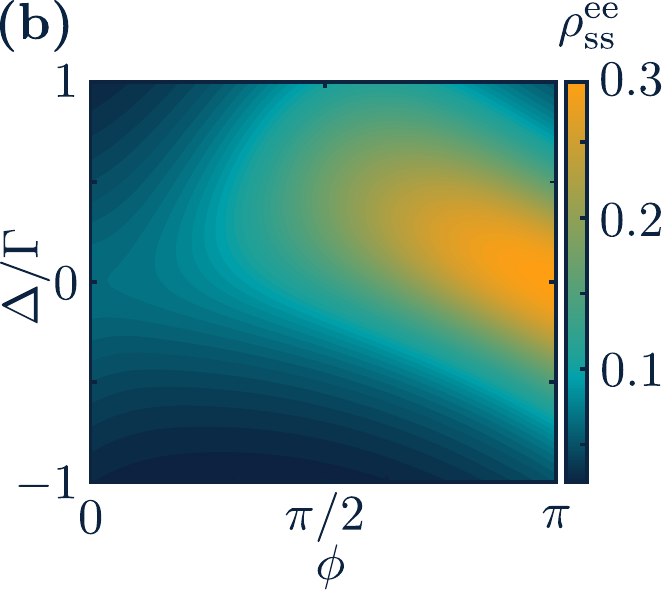}%
}
\caption{ Excited state probability of the atom in front of the mirror: (a) Probability dynamics for the various driving strengths $\Omega=0.1,\;0.2,\;0.5\Gamma$, delay time $\Gamma\tau=20$, $\phi=\pi$, $\chi=80$, $\Delta/\Gamma=0$. (b) Excited state probability in the steady state as a function of the round-trip phase $\phi$ and the detuning $\Delta/\Gamma=0$, $\Omega/\Gamma=0.5$,  $\chi=50$. The Trotter step is $\Gamma\Delta t=0.1$ for both plots.}
\label{fig:mirrorplot}
\end{figure}

\subsubsection{Atomic dynamics and steady state}
In this subsection we present results obtained from solving for the dynamics and the steady state of the atom in front of the mirror, using the methods developed in the previous sections. 

In the figure Fig.~\ref{fig:1atom} we plot the evolution of the atomic excitation probability for a resonant driving field as a function of time, $\rho_{ee}(t)=\bra{e}\rho(t)\ket{e}$, for up to 15 round-trip times with long time delays $\Gamma\tau=20$, and a round-trip phase of $\phi=\pi$. To interpret the results, it is useful to recall that a two-level atom in its ground state acts like a mirror for photons in a frequency band of width $\Gamma$ around the two-level transition frequency. With the choice of $\phi=\pi$, the delay line and the atom therefore form a perfect cavity for a (single) photon that is resonant with the atomic transition frequency. This effect leads to a dynamical accumulation of photons in the delay line, as long as the atomic excitation probability is small. This dynamic proceeds until the field in the delay line is strong enough to effectively saturate the two-level atom, rendering it non-reflective and allowing photons to leak out of the delay line. This interplay between photon trapping and atom saturation determines the steady state. If the coherent driving field is very weak, it takes several round-trip times until this point is reached, while for a stronger drive the atom saturates much quicker due to the coherent drive. 

If the coherent driving field is not resonant with the two-level system transition frequency, i.e.,if  the detuning $\Delta$ is non-zero, the reflectivity of the atom and thus the trapping capabilities of the setup change. In fact, this trapping capabilities are determined by a non-trivial interplay between the detuning and the round-trip phase. This is displayed in Fig.~\ref{fig:rhoee}, where we show the steady state excitation  probability of the atom, which is related to the photon number in the delay line via the input-output relation.

\begin{figure}
   \subfloat[\label{fig:tss0}]{%
  \includegraphics[width=0.5\columnwidth]{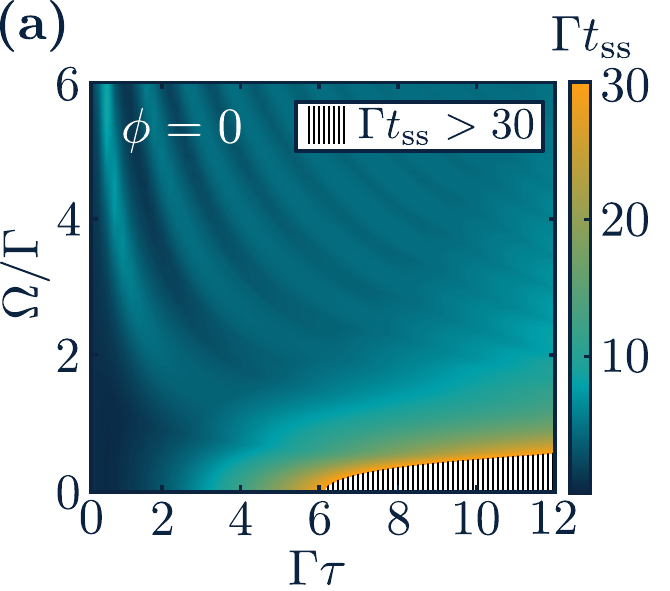}%
}\hfill
\subfloat[\label{fig:tssphi}]{%
  \includegraphics[width=0.5\columnwidth]{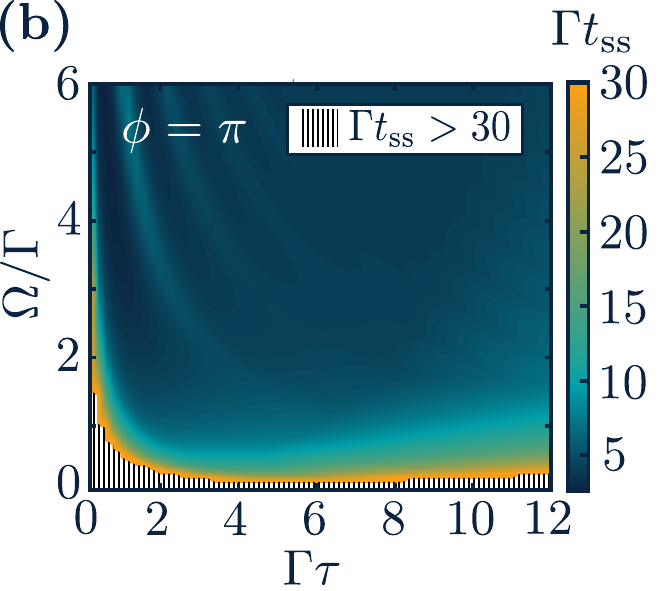}%
}
\vspace{-0.5cm}
\subfloat[\label{fig:tsssmall0}]{%
  \includegraphics[width=0.5\columnwidth]{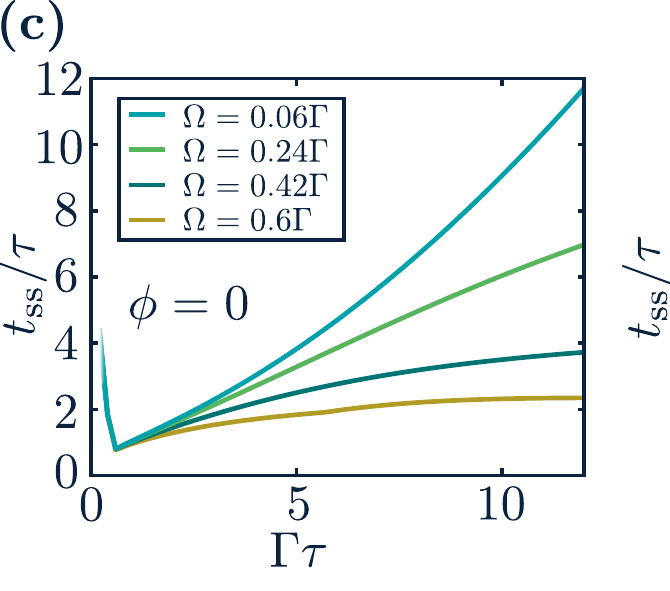}%
}
\hfill
   \subfloat[\label{fig:tsssmallphi}]{%
  \includegraphics[width=0.5\columnwidth]{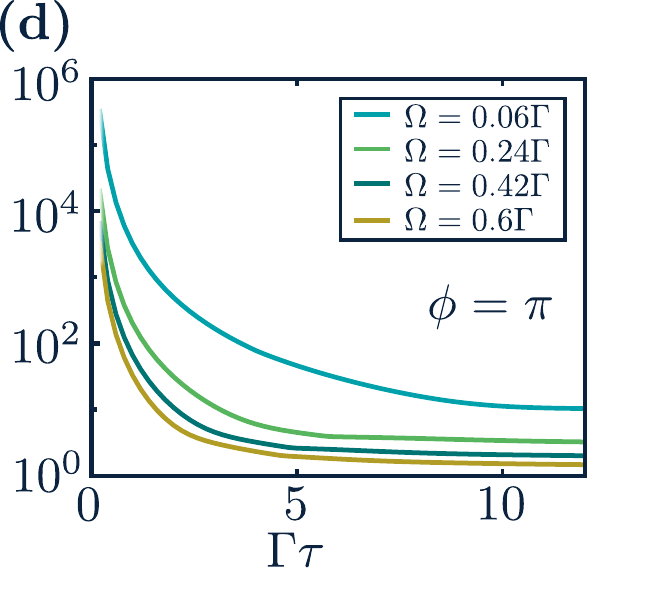}%
}
\vspace{-0.5cm}
\subfloat[\label{fig:tssphiom02}]{%
  \includegraphics[width=0.5\columnwidth]{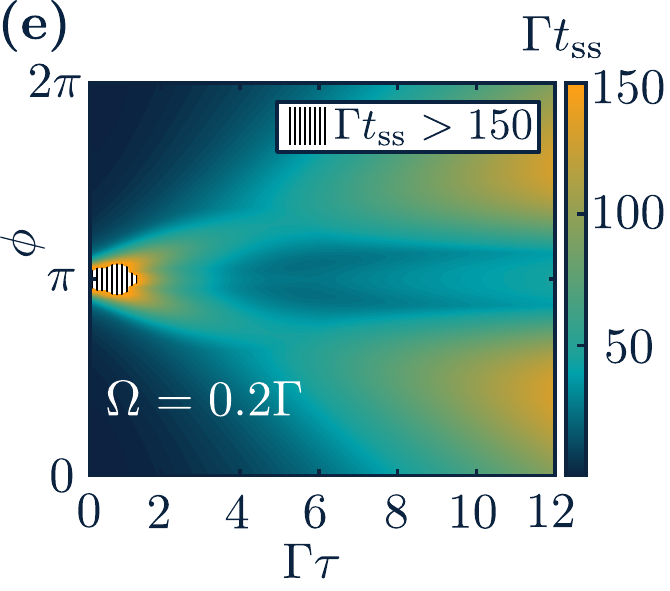}%
}
\hfill
\subfloat[\label{fig:tsstrilobyte}]{%
  \includegraphics[width=0.5\columnwidth]{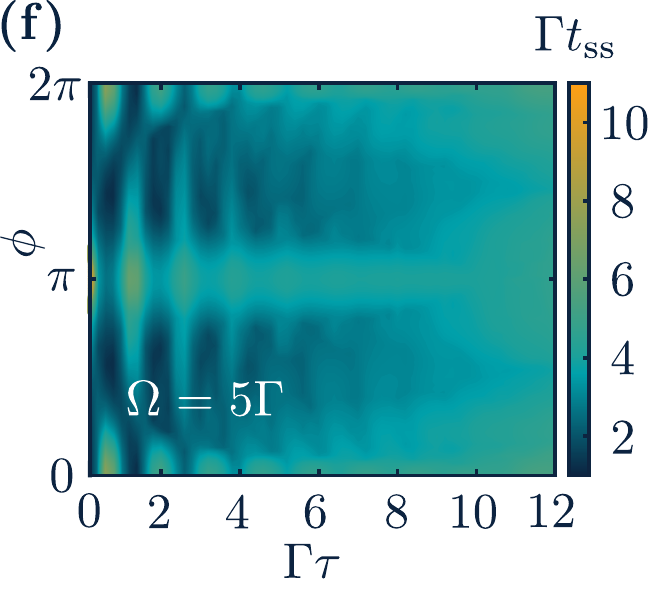}%
}
\caption{Steady state time $t_{\rm ss}$ of the atom in front of the mirror: (a--b) The dependence on Rabi frequencies and delay times, $\phi=0$, and $\phi=\pi$. (c--d) The steady state time relative to the delay time in the small driving limit for different values of $\Omega$, $\phi=0$ and $\phi=\pi$. (e--f) The steady state time calculated for the different phases and delay times, $\Omega=0.2\Gamma$ and $\Omega=5\Gamma$. For all plots the parameters are: $\Gamma\Delta t=0.1$, $\chi=250$, $\Delta/\Gamma=0$.}
\label{fig:tss}
\end{figure}

As it was noted, our method allows us also to directly access the time it takes the system to relax to its steady state, $t_{\rm ss}$. Figure~\ref{fig:tss} shows $t_{\rm ss}$ for a resonant drive as a function of  delay time and driving strength as well as round-trip phase. Due to the photon-trapping mechanism discussed earlier, we observe long relaxation times in the regime of weak driving and long delay times, for large enough Rabi frequencies $t_{\rm ss}$ oscillates with a period proportional to $\frac{2\pi}{\Omega}$. This can be understood by noting that the Rabi oscillations of the atom affect the probability of the photon to be reflected by the atom. When the delay time becomes large enough for the atom to reach an equilibrium state during a round-trip time, these oscillations damp out. The round-trip phase also affects the chances of a photon to be trapped, thus increasing the steady state time as it is seen in Fig.~\ref{fig:tssphi}. 

\subsubsection{Output field properties}
The infinite chain algorithm for the atom in front of the mirror discussed earlier can be used to calculate the steady state properties of the output field, such as the spectrum and the intensity correlation functions.

\begin{figure}
   \subfloat[\label{fig:Spectrum20}]{%
  \includegraphics[width=0.5\columnwidth]{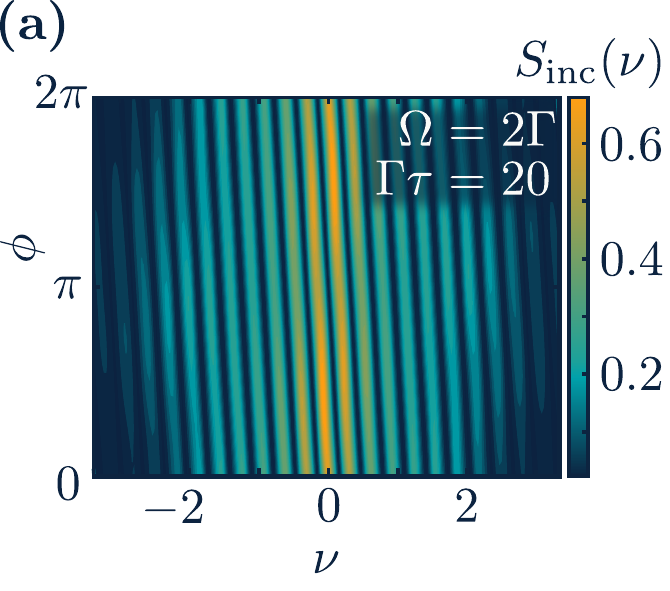}%
}\hfill
\subfloat[\label{fig:G2}]{%
  \includegraphics[width=0.5\columnwidth]{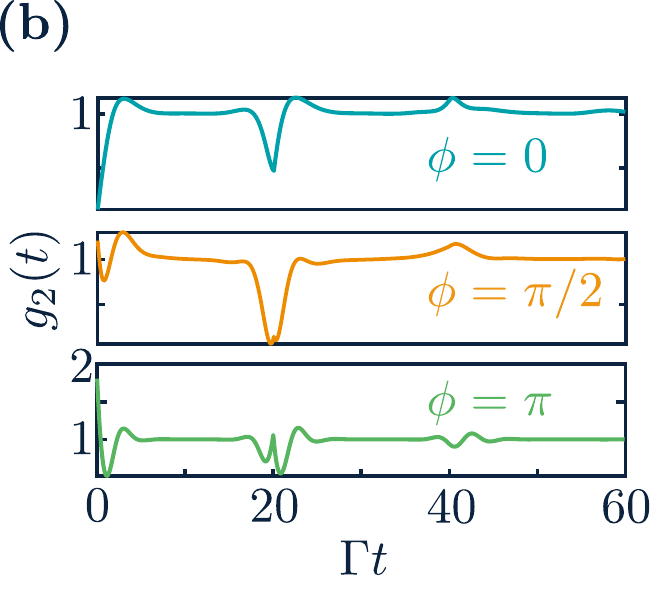}%
}
\caption{Steady state output field properties for the setup with the atom in front of the mirror:  (a) The dependence of the incoherent part of the spectrum $S_{\rm inc}(\nu)$ on the round-trip phase. (b) The dependence of the second-order correlation function $g_2(t)$ on the delay time $\Gamma\tau$ for the round-trip phases $\phi=0,\pi/2,\pi$; for both plots $\Omega=2\Gamma,\;\Gamma\tau=20,\;\Delta/\Gamma=0$, $\chi=80$, $\Gamma\Delta t=0.01$.}
\end{figure}

 The steady state spectrum of the output field detected at the open side of the waveguide is given by
\begin{equation}
S(\nu)=2\Re\left\{\int_0^\infty dt'\left<b_{\rm out}^\dagger(t)b_{\rm out}(t-t')\right>e^{-i\nu t'}\right\},
\end{equation}
where the output field operator is obtained using the input-output formalism as
\begin{equation}\label{eqn:inout}
    b_{\text{out}}(t)=\sqrt{\gamma_L}c_L(t)+\sqrt{\gamma_R}e^{i\phi}c_R(t-\tau)+e^{i\phi}b(t-\tau).
\end{equation}
Based on this expression, the output spectrum can be obtained from two-times system correlation functions (see Appendix \ref{app:infchain} for details). The incoherent part of the spectrum in the case of the long delay time and for different round-trip phases is shown in Fig.~\ref{fig:Spectrum20} and exhibits a pattern of minima and maxima with a periodicity proportional to $1/\tau$. This periodicity is a result of the correlations between the photons emitted with the time difference $\tau$. Using the input-output formalism Eq.~\eqref{eqn:inout}, one can also calculate the normalised second-order correlation functions of the output field
\begin{equation}
g_2(t')=\frac{\left<b_{\rm out}^\dagger(t)b_{\rm out}^\dagger(t-t')b_{\rm out}(t-t')b_{\rm out}(t)\right>}{\left<b_{\rm out}^\dagger(t)b_{\rm out}(t)\right>\left<b_{\rm out}^\dagger(t-t')b_{\rm out}(t-t')\right>}
\end{equation}
The result is shown in Fig.~\ref{fig:G2} for different round-trip phases and exhibits both bunching and antibunching behavior depending on the round-trip phase.

   \begin{figure}[t]
   \subfloat[\label{fig:2atoms}]{%
  \includegraphics[width=0.5\columnwidth]{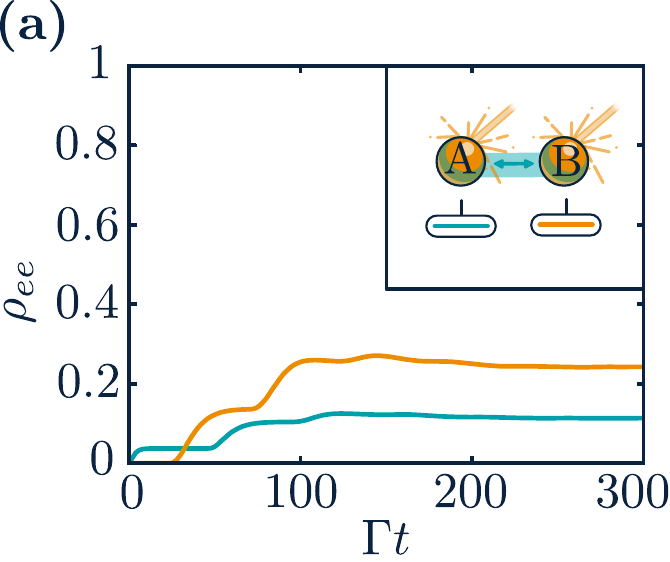}%
}\hfill
\subfloat[\label{fig:3atoms}]{%
  \includegraphics[width=0.5\columnwidth]{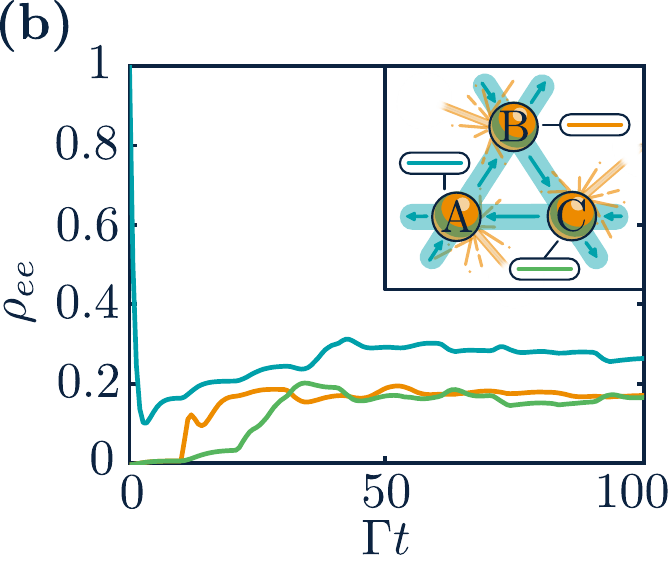}%
}
\caption{Excited state probabilities dynamics for the multi-node setups: (a) $\rho^{A}_{\rm ee}$ and  $\rho^{B}_{\rm ee}$ of the two atoms A and B connected through the bidirectional waveguide as a function of $\Gamma t$. Both atoms are initially in the ground state, with parameters $\gamma_1=\Gamma/2$, $\gamma_2=\Gamma/10$, $\Omega_1=\Gamma/5$, $\Omega_2=0$, $\phi=\pi$, $\Gamma\tau=20$, $\chi=50$. (b) The excited state probabilities $\rho^{A}_{\rm ee}$, $\rho^{B}_{\rm ee}$, and $\rho^{C}_{\rm ee}$ of the three atoms A, B, and C connected through 1D unidirectional waveguides as a function of $\Gamma t$, initially the atoms are in the excited, ground and ground states; $\gamma_1=\gamma_2=\gamma_3=\Gamma/2$, $\Omega_1=\Gamma/2$, $\Omega_2=\Omega_3=\Gamma/10$, $\phi=\pi$, $\Gamma\tau=10$, $\chi=25$. For both plots $\Delta/\Gamma=0$ for all nodes and $\Delta t=0.1$} 
\label{fig:manyatomsplot}
\end{figure}

\subsection{Other networks}
In this subsection we present results from the application of our method to other simple quantum optical networks, connecting two or three nodes. Fig.~\ref{fig:manyatomsplot}(a) shows the dynamics of a pair of two-level atoms coupled to a bidirectional waveguide at two distant points. The time delay is a result of the propagation time a photon needs to travel between the two systems. For this case we assume $\gamma_{i,R}=\gamma_{i,L}=\gamma_i$ and $\Gamma=2\gamma_1$. Next, Fig.~\ref{fig:manyatomsplot}(b) shows the dynamics of three nodes connected pairwise with unidirectional waveguides, with equal time delay in each interconnect. Here, again, $\gamma_{i,R}=\gamma_{i,L}=\gamma_i$ and $\Gamma=2\gamma_1$. These results can be obtained through an adaptation of the derivation given in Sec.~\ref{sec:1Dcascchain} and a corresponding, simple modification of the algorithm given in Sec.~\ref{sec:nuemricalMethods}. We discuss these generalizations to more complicated networks in detail in Appendix~\ref{app:multiatom}.  
\section{Mean-field approximation}\label{sect:meanfield}
In this section we will use the insights of the numerical results from Sec.~\ref{sec:results} to propose a semi-analytical solution based on the mean-field approximation of the 1D cascaded chain. As Fig.~\ref{fig:entropy} indicates, the correlations in the propagator $E\strut^{[m]}(s)$ are small when both the time delay and the Rabi frequency are large. This suggests that in this regime the $m$-site propagator $E\strut^{[m]}(s)$ can be approximated as a tensor product of local propagators \begin{align}E\strut^{[m]}\!(s)\approx\bigotimes_{j=1}^m E^{\rm mf}_j(s)
,\end{align} 
where $E^{\rm mf}_j(s)$ is a mean-field propagator at site $j$. We use standard  mean-field approach to determine these local propagators, starting with the equation for the total propagator $E\strut^{[m]}\!(s)$, Eq.~\eqref{eq:Prop_differential}: Assuming  the above product form of the propagator, one readily obtains the equation of motion for the local mean-field propagator $E_i^{\rm mf}(s)$ by tracing out all sites except $i$ in Eq.~\eqref{eq:Prop_differential}. This procedure gives
\begin{equation}\label{eqn:dedsmf}
\frac{d}{ds}E_i^{\rm mf}\!(s)=\mc{L}^{\rm mf}_i(s) E^{\rm mf}_i(s),
\end{equation}
with initial condition $E_i^{\rm mf}(0)=\mathbb{1}$. Here $\mc{L}^{\rm mf}_i(s)$ is a mean-field Lindblad operator at site $i$, which is given via 
\begin{equation}\label{eqn:dedsmffull}
\mc{L}^{\rm mf}_i(s) E^{\rm mf}_i(s)={\rm tr}_{i\pm 1}\left\{(\mc{L}_{i-1,i}^{\rm casc}+\mc{L}_{i,i+1}^{\rm casc})\bigotimes_{j=i-1}^{i+1}
E_j^{\rm mf}\!(s)
\right\}.
\end{equation}
Here ${\rm tr}_{i\pm 1}$ denotes the partial trace over  sites $i-1$ and $i+1$. 
Straightforward algebra allows one to bring the above expression into a particular transparent form. 
\begin{align}\label{eqn:loclinbl}
\mc{L}^{\rm mf}_i(s)E_{i}^{\rm mf}(s)&=\mc{L}^{\rm M}_iE_{i}^{\rm mf}(s)-\tfrac{i}{\hbar}[h_i(s), E_{i}^{\rm mf}(s)]. 
\end{align}
Here we introduced the notation
$\mc{L}_i^MX=-\frac{i}{\hbar}\left[H_{{\rm sys}, i},\right]+\mc{D}\left[L_i\right]+\mc{D}\left[R_i\right]$
as well as
$h_i(s)=i\hbar(r_{i-1}^\ast(s) L_i-r_{i-1}(s) L_i^\dag)$. In the last expression we used the shorthand notation for the maps $r_{i-1}(s)X_{i-1}={\rm tr}_{i-1}\{R_{i-1} E_{i-1}^{\rm mf}(s)X_{i-1}\}$  and $r^\ast_{i-1}(s)X_{i-1}={\rm tr}_{i-1}\{ E_{i-1}^{\rm mf}(s)X_{i-1}R^\dag_{i-1}\}$, which map  operators in the Hilbert space of replica $i-1$ (specifically, density matrices) to a  c-number. Note that $\mc{L}_i^{\rm mf}$ is manifestly of Lindblad form. The first term in Eq.~\eqref{eqn:loclinbl}, $\mc{L}_i^M$, is in fact simply the generator of a Markovian master equation describing the replica system $i$ coupled to a bath \textit{without} time-delayed feedback, such as a waveguide that is open on both ends. The second term captures the effect of the time-delayed feedback on a mean field level: It generates an additional coherent evolution of the replica $i$, dependent on replica $i-1$. Specifically, one can interpret this second term as an additional coherent field driving the replica system $i$. The amplitude of this driving field is simply determined by the expectation value of the output field of the neighboring replica at site $i-1$. It is this second term that renders the mean-field equations non-linear. Note that the mean-field equations for $E_i^{\rm mf}(s)$ depend only on $E_{i-1}^{\rm mf}(s)$ but not on $E_{i+1}^{\rm mf}(s)$. This is the direct consequence of the unidirectional nature of the cascaded chain. 

   \begin{figure}
   \subfloat[\label{fig:mffidelityphi0}]{%
  \includegraphics[width=0.5\columnwidth]{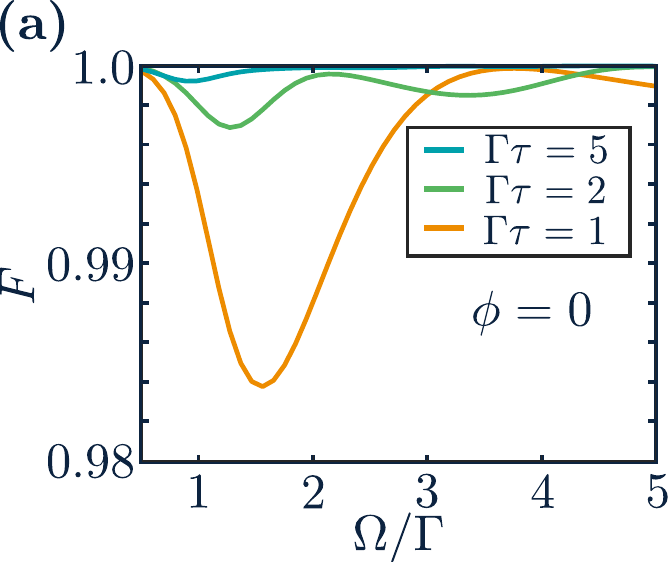}%
}
\subfloat[\label{fig:mffidelityphipi}]{%
  \includegraphics[width=0.5\columnwidth]{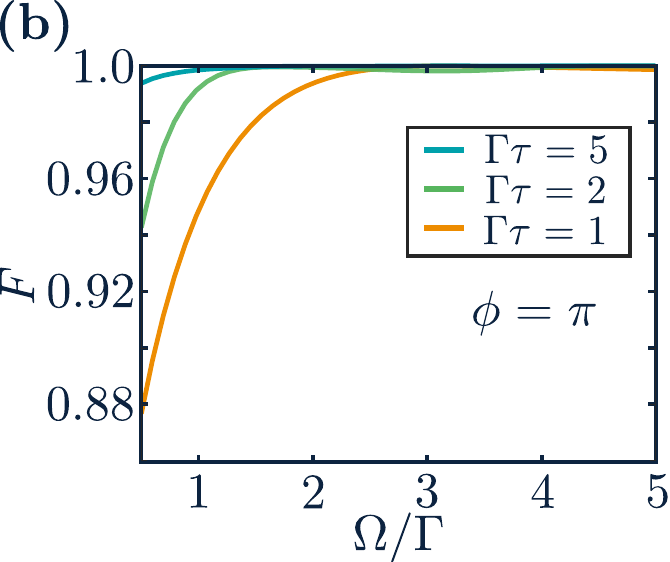}%
}\vspace{-0.5cm}
   \subfloat[\label{fig:mfspectrum55}]{%
  \includegraphics[width=0.5\columnwidth]{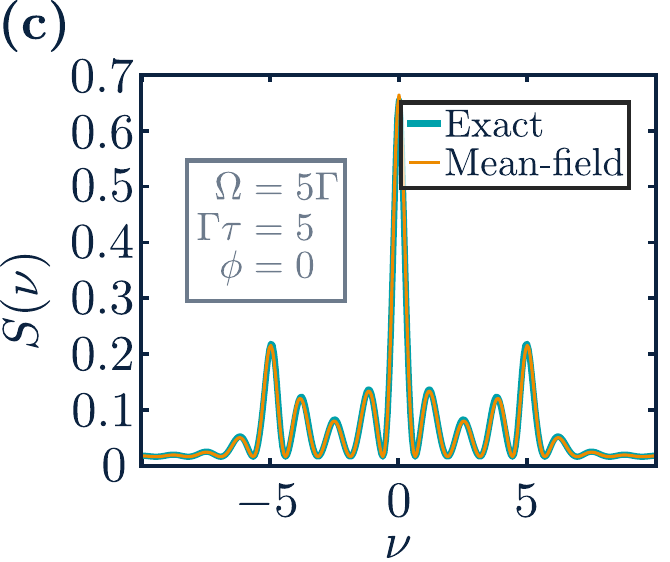}%
}
\subfloat[\label{fig:mfspectrum52}]{%
  \includegraphics[width=0.5\columnwidth]{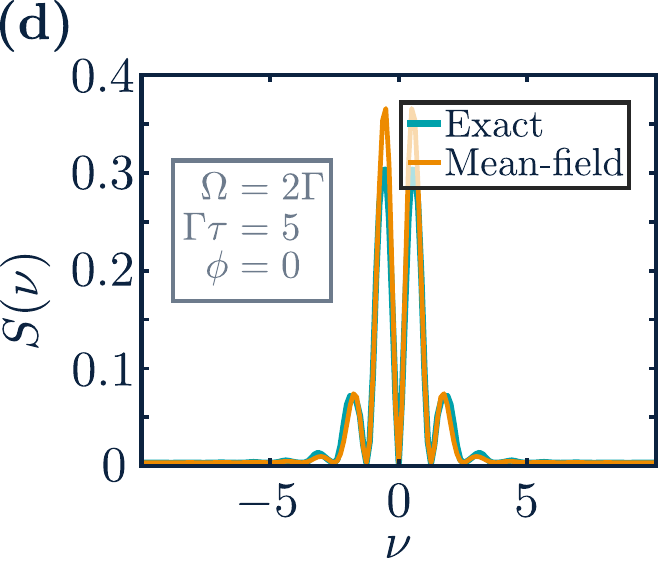}%
}\vspace{-0.5cm}
   \subfloat[\label{fig:mfspectrum15}]{%
  \includegraphics[width=0.5\columnwidth]{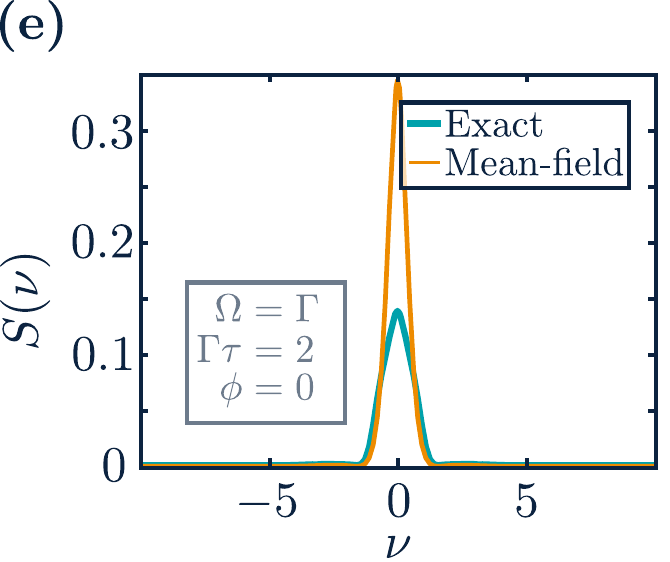}%
}
\subfloat[\label{fig:mfspectrum11}]{%
  \includegraphics[width=0.5\columnwidth]{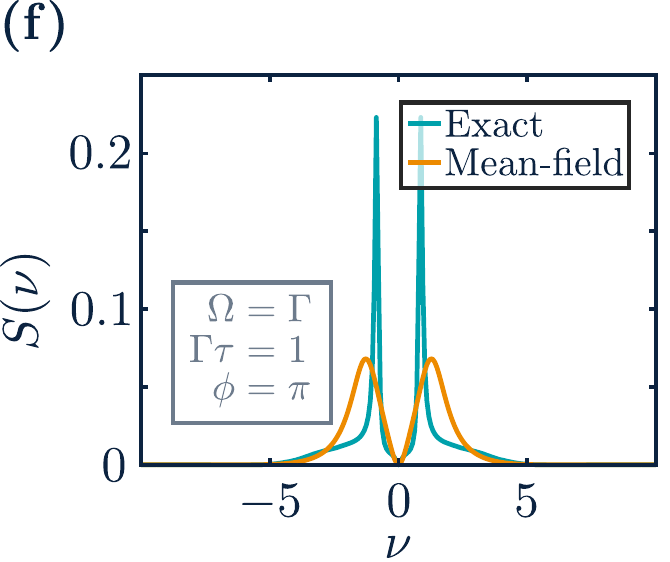}%
  }
\caption{Comparison of the exact steady state solution and the mean-field approximation for the atom in front of the mirror: (a-b) The fidelity of the steady state mean-field solution with the exact solution, $F=\left({\rm tr}\sqrt{\sqrt{\rho_{\rm ss}}\rho_{\rm ss}^{\rm mf}\sqrt{\rho_{\rm ss}}}\right)^2$, dependence on the driving with the delay times $\Gamma\tau=1,\;2,\;5$, (a) $\phi=0$ and (b) $\phi=\pi$. Panels (c)-(f) show a comparison between the exact output field spectrum and the mean-field approximation. The parameters are: (c) $\Omega/\Gamma=5$, $\Gamma\tau=5$; (d) $\Omega/\Gamma=2$, $\Gamma\tau=5$; (e) $\Omega/\Gamma=1$, $\Gamma\tau=2$; (f) $\Omega/\Gamma=1$, $\Gamma\tau=1$. Each spectrum is calculated for $\phi=\pi$. In all panels we used  $\Delta/\Gamma=0$, $\chi=50$ and $\Gamma\Delta t=0.01$} 
\label{fig:mf}
\end{figure}

With the expression \eqref{eqn:loclinbl} we can solve the non-linear, coupled mean-field equations \eqref{eqn:dedsmf} to obtain the propagator $E^{\rm mf}(s)$ in mean-field approximation and consequently calculate from it mean-field approximation of all quantities of interest as discussed in the previous Sections. 
 Fig.~\ref{fig:mffidelityphi0} shows the fidelity \cite{doi:10.1080/09500349414552171} between the steady state in mean-field approximation and the exact steady state  calculated in the previous section. As expected, the mean-field approximation improves and approaches 1 when both the delay time and the driving increase. Remarkably, the mean-field approximation can also capture relevant two-time correlation functions successfully: Fig.~\ref{fig:mf}(b)-(d) show the incoherent part of the output field spectrum calculated both by the exact algorithm and using the mean-field approximation for different parameters. Again, as the driving strength and the delay time increase, the mean-field approximation becomes more accurate. In the case of sufficiently large Rabi frequencies and delay times, the mean-field picture allows for a simple interpretation of the spectrum: It is given by the standard Mollow triplet found in the output of a strongly driven two-level system coupled to a Markovian bath \cite{PhysRev.188.1969}, which is modulated with a frequency $1/\tau$ as a result of constructive (destructive) interference between the emitted photons and those returning to the atom from the delay line (previous replica).

We conclude this section by noting that the mean-field approximation is applicable also in more general quantum optical setups with time delays. In particular, we expect this method to be useful in situations where the mapping from the non-Markovian system of interest to a corresponding Markovian many-body system, following the method outlined in Sec.~\ref{sec:mapping1Dchain}, results in a Markovian description in more than 1D. 

\section{Conclusion}
In this work we developed a novel approach for solving problems with continuous coherent quantum feedback involving time delays. Our method is numerically exact, and we demonstrated its efficiency for several examples. One of the most interesting challenges in going beyond the models presented in this work is to understand if it is possible to construct examples where the methods developed here fail. For instance, this could happen if one found examples of 1D cascaded chains whose propagators are not rapidly mixing, such that the corresponding operator entanglement does not obey  an area law. Identifying such setups would potentially allow to engineer quantum optical setups that can produce qualitatively more complex output states, such as states with algebraically decaying correlation functions. This would have important implications for photonic quantum simulation approaches \cite{barrettSimulatingQuantumFields2013,PhysRevX.5.041044}.

In this work we focused our analysis on systems where all time delays are equal, which allowed us to map the problem to a Markovian problem in one dimension and in turn solve it via MPS techniques. More complex networks with multiple, incommensurate time delays map to Markovian many-body problems in more than one dimension. We expect that this is a regime where the mean field approach developed in this work could be especially useful. 
\section*{Acknowledgements}

We thank Peter Zoller, Helmut Ritsch and Crispin Gardiner for helpful discussions. We acknowledge financial support from the ERC Starting grant QARA (grant no. 101041435), the European Unions Horizon 2020 research
and innovation program under Grant Agreement No.
101079862 (PASQuanS2), and by the EU-QUANTERA project TNiSQ
(N-6001). The computational results presented have been achieved (in part) using the HPC infrastructure LEO of the University of Innsbruck.



\section*{Appendix}
\appendix
\subsection{Mapping to 1D cascaded chain}\label{app:1Dchain}
We will now prove the direct correspondence between the equation for the reduced density matrix of the atom in front of the mirror and the 1D cascaded chain. This proof can be naturally generalized to the case of multi-node networks; however, for illustrational purposes we consider here the simplest example. We start out with the equation \eqref{eqn:rhoTN} and proceed by decomposing the superoperator $T_i$ in two superoperators $T^R_i$ and $T^L_i$, with the first one acting both on the system and the time bin $i+k$ and the second one on the system and the time bin $i$ as in Fig.~\ref{fig:densitymatrixsandwich}(b), $ T_i X=T^R_iT^L_iX,$
where we define $T^R_iX=V_i^RX{V_i^R}^\dag$ and 
$T^L_iX=\textrm{tr}_{\mc{H}_i}\{V_i^LX{V_i^L}^\dag\}$ with the unitary $V_{i}^L=\exp(-\frac{i}{\hbar}H_{\text{sys}}\frac{\Delta t}{2}+\Upsilon_i^L)$, and the isometry $V_{i}^R=\exp(-\frac{i}{\hbar}H_{\text{sys}}\frac{\Delta t}{2}+\Upsilon_i^R)\ket{0}_{i+k}$ correspondingly. This decomposition is correct up to Trotter errors that vanish in the $\Delta t\rightarrow 0$ limit. Note that (up to higher-order Trotter terms)  this decomposition is symmetric, $ T_i X=T^R_iT^L_iX=T^L_iT^R_iX$.

 We rewrite the equation \eqref{eqn:rhoTN} using the unitaries and the isometries introduced above as
\begin{widetext}
\begin{align}\label{eqn:rhoVNdet}
\rho_{\rm sys}(t_n)=\textrm{tr}_{\mc{H}_{B}}\{{V^R_{n-1}}{V^L_{n-1}}\dots {V^R_{i}}{V^L_{i}}\dots {V^R_{0}} {V^L_{0}}\; \rho_{\rm sys}(t_0)\bigotimes_{i=0}^{k-1}\ket{0_i}\bra{0_i} V^{L\dag}_{0} V^{R\dag}_{0} \dots V^{L\dag}_{i}V^{R\dag}_{i}\dots V^{L\dag}_{n-1}V^{R\dag}_{n-1}\}.
\end{align} 
We will now carefully work out the whole expression by taking the partial trace over the bath degrees of freedom of each time bin. To do so, we introduce matrix elements of the operators $V_i^L$ and $V_i^R$ as
$V^L_{i,b,a}=\bra{b}V^L_{i}\ket{a}$ and $V^R_{i,a,b}=\bra{a}V^R_{i}\ket{b}$, where $\ket{a}$ and $\ket{b}$ are the basis states of the $d$-dimensional system Hilbert space. Note that these matrix elements act on the  photonic time bins. Specifically, $V^L_{i,b,a}$ is an operator that acts on time bin $i$, and $V^R_{i,a,b}$ is a state of the time bin $i+k$. 
We now can write the matrix elements of the system density matrix at time $t_n$
\begin{align}\label{eqn:rhoVNdetindex}
\rho^{\rm sys}_{a_n,a'_{n}}(t_n)&=\sum_{{\bf a},{\bf b},{\bf a'},{\bf b'}}\textrm{tr}_{\mc{H}_{B}}\{V^{R}_{n-1,a_n,b_{n-1}}V^{L}_{n-1,b_{n-1},a_{n-1}}\dots{V^{R}_{i,a_{i+1},b_{i}}V^{L}_{i,b_{i},a_{i}}}\dots \\\nonumber
&\dots V^{R}_{0,a_{1},b_{0}}{V^{L}_{0,b_{0},a_{0}}}\rho_{a_{0},a'_{0}}(t_0)\bigotimes_{i=0}^{k-1}\ket{0_i}\bra{0_i} 
V^{L\;\dag}_{0,a'_{0},b'_{0}}V^{R\;\dag}_{0,b'_{0},a'_{1}} \dots V^{L\;\dag}_{k,a'_{i},b'_{i}}V^{R\;\dag}_{i,b'_{i},a'_{i+1}}\dots V^{L\;\dag}_{n-1,a'_{n-1},b'_{n-1}}V^{R\;\dag}_{n-1,b'_{n-1},a'_{n}}\},  
\end{align}
where the sum goes over the indexes ${\bf a}=a_0,\dots,a_{n-1},\;{\bf b}=b_0,\dots,b_{n-1},\;{\bf a'}=a'_0,\dots,a'_{n-1},\;{\bf b'}=b'_0,\dots,b'_{n-1}$ and where we used $\rho_{a_{0},a'_{0}}(t_0)=\bra{a_0}\rho(t_0)\ket{a'_0}$. To perform the partial trace over the time bins, we rearrange the terms in the above expression, grouping together all the terms that involve the same time bin. Therefore, we group pairs $V^R_{i-k,a_{i-k+1},b_{i-k}}$ and $V^L_{i,b_{i},a_{i}}$ (and analogously $V^{R\;\dag}_{i-k,b'_{i-k},a'_{i-k+1}}$ with $V^{L\;\dag}_{i,a'_{i},b'_{i}}$), since these are the only terms involving the time bin $i$. This allows us to trace out all time bins $i$ that appear in the state (i.e., the time bins from $i=0$ to $i=n+k-1$), sequentially, which gives rise to three different types of terms. The first type of terms arises from the trace over the time bins $i=0,\dots,k-1$, which gives terms of the form 
\begin{align}\label{eqn:Wbdl}
W^{(L)}_{b_i,b_i',a_i,a_i'}=\textrm{tr}_{\mc{H}_{i}}\{{V^{L}_{i,b_{i},a_{i}}} \ket{0_i}\bra{0_i}V^{L\;\dag}_{i,a'_{i},b'_{i}}\}.
\end{align}
The second type of terms arise from the trace over the time bins $i=k,\dots, n-1$, which gives terms of the form  
\begin{align}\label{eqn:Wcascmap}
W^{(LR)}_{b_{i},b'_{i},a_{i-k+1},a'_{i-k+1},a_i,a_i',b_{i-k},b'_{i-k}}=\textrm{tr}_{\mc{H}_{i}}\{{V^{L}_{i,b_{i},a_{i}}} {V^{R}_{i-k,a_{i-k+1},b_{i-k}}}V^{R\;\dag}_{i-k,b'_{i-k},a'_{i-k+1}}  V^{L\;\dag}_{i,a'_{i},b'_{i}}\}.
\end{align}
The third type of terms are obtained from the trace over the time bins $i=n,\dots, n-1+k$, which gives terms of the form 
\begin{align}\label{eqn:Wbdr}
W^{(R)}_{a_{i-k+1},a_{i-k+1}',b_{i-k},b'_{i-k}}=\textrm{tr}_{\mc{H}_{i}}\{{V^{R}_{i-k,a_{i-k+1},b_{i-k}}} V^{R\;\dag}_{i-k,b'_{i-k},a'_{i-k+1}}\}.
\end{align}
\begin{figure}
             \includegraphics[width=0.8\linewidth]{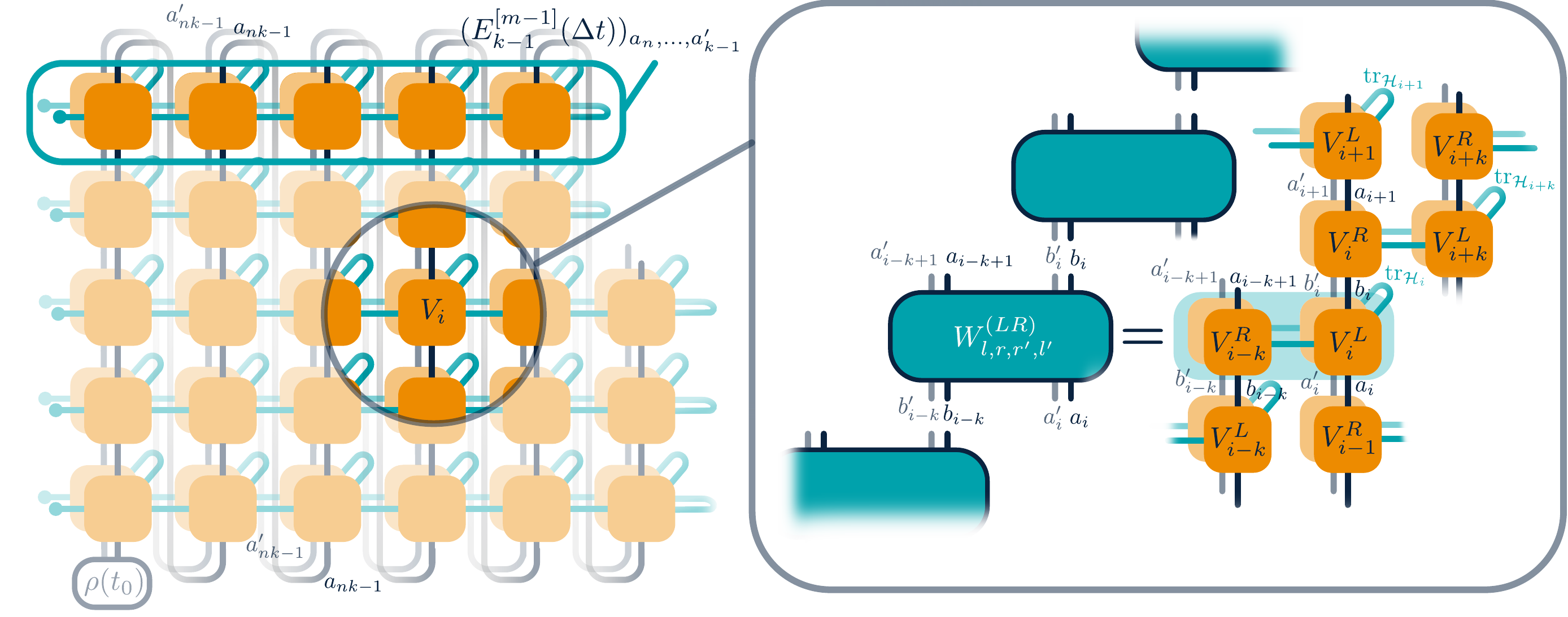}
				\caption{Mapping to 1D cascaded chain: Reduced density matrix of the system $\rho_{\rm sys}(t_n)$ as a tensor network, the connection of the isometry $V_i$ to its neighbors is encircled and shown in details to the right, where we separate $V_i^R$ and $V_i^L$ and explicitly write the indexes associated with each leg. Traces over the bath degrees of freedom are indicated as the contracted blue legs. Contraction of $V_{i-k}^R$ and $V_i^L$ creates local propagator $W^{(LR)}$ leading to a ladder-like structure made of these propagators for each horizontal layer in the original tensor network. Note that in the main text we used the fact that Trotter decomposition is symmetric $ T_i X=T^R_iT^L_iX=T^L_iT^R_iX$ and the resulting chain of propagators has a different (zigzag) structure. In a light blue rectangle, we enclose the transfer operator $E^{[m-1]}_{k-1}(\Delta t)$ and denote the indexes through which it connects to the lowest transfer operator $E_0^{[m]}(\Delta t)$.}
    \label{fig:indexes}
			\end{figure}

To proceed, we evaluate now all three of these terms. To do so, we use the definition of $V^L$ and $V^R$ and expand to the first order in $\Delta t$ (recalling that the Ito increment $\Delta B_i$ gives contributions in order $\sqrt{\Delta t}$)
 \begin{align}\label{eqn:VLdt}
V^L_{i,b,a}&=\delta_{b,a}-\frac{i}{2\hbar}H^{\text{sys}}_{b,a}\Delta t+(L_{b,a}\Delta B_{i}^\dagger-{\rm h.c.})+\\\nonumber
&+
\frac{1}{2}\left((L^2)_{b,a}\Delta B_{i}^{\dagger 2} -(LL^{\dagger})_{b,a}\Delta B_{i}^{\dagger}\Delta B_{i}-(L^{\dagger}L)_{b,a}\Delta B_{i}\Delta B_{i}^{\dagger}+(L^{\dag 2})_{b,a}\Delta B_{i}^{2}\right)+\ldots\\
V^R_{i,a,b}&=\lr{\delta_{a,b}-\frac{i}{2\hbar}H^{\text{sys}}_{a,b}\Delta t+R_{a,b}\Delta B_{i+k}^\dagger
-(R^{\dagger}R)_{a,b}\Delta t
}\ket{0_{i+k}}+\frac{1}{2}(R^2)_{a,b}\Delta B_{i+k}^{\dagger 2}\ket{0_{i+k}}+\ldots\label{eqn:VRdt}
\end{align}
With this one can evaluate the above expressions (\eqref{eqn:Wbdl}-\eqref{eqn:Wbdr}) using the commutation relation of the Ito increment operator, $[\Delta B^\dagger_i,\Delta B_j]=\Delta t\delta_{ij}$ . For the term \eqref{eqn:Wbdl} we obtain to the first order in $\Delta t$
\begin{align}
W^{(L)}_{b,b',a,a'}&=\delta_{b,a}\delta_{b',a'}-\frac{i}{2\hbar}(H_{\rm sys})_{b,a}\delta_{a',b'}\Delta t+\frac{i}{2\hbar}\delta_{b,a}(H_{\rm sys})_{a',b'}\Delta t+\frac{1}{2}[2L^\dag_{b,a}L_{a',b'}-(L^\dag L)_{b,a}\delta_{a',b'}-\delta_{b,a}(L^\dag L)_{a',b'}]\Delta t
\end{align}
The right-hand side can be identified with the propagator generated by a Lindblad operator $\mc{L}^{\textrm{bd} L}$ given in Eq.~\eqref{eqn:boundaryterms} of the main text and written here explicitly in the basis of the system Hilbert space. To leading order in $\Delta t$, this can be rewritten as
\begin{align}
W^{(L)}_{b,b',a,a'}&=(e^{\Delta t \mc{L}^{\textrm{bd} L}})_{b,b',a,a'},
\end{align}
where we define the matrix element of a superoperator as $(e^{\Delta t \mc{L}^{\textrm{bd} L}})_{b,b',a,a'}={\rm tr}_{\rm sys}\{\ket{b}\bra{b'}(e^{\Delta t \mc{L}^{\textrm{bd} L}})\ket{a}\bra{a'}\}$.
Analogously, we find for the third term \eqref{eqn:Wbdr} 
\begin{align}
W^{(R)}_{a,a',b,b'}=(e^{\Delta t \mc{L}^{\textrm{bd} R}})_{a,a',b,b'},
\end{align}
where $\mc{L}^{\textrm{bd} R}$ is the term given in Eq.~\eqref{eqn:boundaryterms} of the main text.
Finally, the term in Eq.~\eqref{eqn:Wcascmap} can be evaluated in a similar way 

\begin{align}\label{eqn:Wcascmaplong}
W^{(LR)}_{l,r,r',l'}
&=\delta_{l,l'}\delta_{r,r'}\!+\delta_{l,l'}O_{r,r'}^{{\rm sys}}+O_{l,l'}^{{\rm sys}}\delta_{r,r'}+\delta_{l,l'}O_{r,r'}^{R}\!\!+O_{l,l'}^{L}\delta_{r,r'}\!+O_{l,r,r',l'}^{LR},
\end{align}
where we introduce superindexes $r=\{a_{i-k+1},b_{i-k}\}$, $l=\{b_{i},a_{i}\}$ and $r'=\{b'_{i-k},a'_{i-k+1}\}$, $l'=\{a'_{i},b'_{i}\}$ (see Fig.~\ref{fig:indexes}).
The total map is a result of the different physical processes: The contributions due to the system Hamiltonian evolution are 
\begin{align*}
O_{r,r'}^{\rm sys}&=-\frac{i}{2\hbar}(H_{\rm sys})_{r}\delta_{r'}-\delta_{r}(H_{\rm sys})_{r'})\Delta t,\\
O_{l,l'}^{\rm sys}&=-\frac{i}{2\hbar}(H_{\rm sys})_{l}\delta_{l'}-\delta_{l}(H_{\rm sys})_{l'})\Delta t,
\end{align*}
the dissipation through the action of the jump operator $R$,
\begin{align*}
O_{r,r'}^{R}=\frac{1}{2}[2R^\dag_{r}R_{r'}-(R^\dag R)_{r}\delta_{r'}-\delta_{r}(R^\dag R)_{r'}]\Delta t,
\end{align*}
the dissipation through the action of the jump operator $L$,
\begin{align*}
O_{l,l'}^{\rm L}=\frac{1}{2}[2L^\dag_{l}L_{l'}-(L^\dag L)_{l}\delta_{l'}-\delta_{l}(L^\dag L)_{l'}]\Delta t.
\end{align*}
The last contribution is a cascaded interaction between two system states
\begin{align*}
O_{l,r,r',l'}^{LR}&=[L^\dag_{l}\delta_{r}R_{r'}\delta_{l'}+\delta_{l}R^\dag_{r}\delta_{r'}L_{l'}-L^\dag_{l}R_{r}\delta_{r'}\delta_{l'}-\delta_{l}\delta_{r}R^\dag_{r'}L_{l'}]\Delta t.
\end{align*}
We note that the total map is then a propagator generated by the cascaded Lindblad operator defined in Eq.~\eqref{eqn:Mastereqn} on two-fold--replicated system Hilbert space
\begin{align}\label{eqn:Wcascmapresult}
W^{(LR)}_{r,l,r',l'}=W^{\rm casc}_{r,l,r',l'}=(\exp(\Delta t\mc{L}^{\rm casc}))_{r,l,r',l'}.
\end{align} 
Using the above derivations, one can trace out all the time bins in the expression \eqref{eqn:rhoVNdetindex} and rewrite it using the three types of propagators we identified earlier
\begin{align}\label{eqn:rhoWindex}
\rho^{\rm sys}_{a_n,a'_{n}}(t_n)&=\sum_{{\bf a},{\bf b},{\bf a'},{\bf b'}}W^{(R)}_{a_{n},a_{n}',b_{n-1},b'_{n-1}}W^{\rm casc}_{b_{n-1},b'_{n-1},a_{n-k},a'_{n-k},a_{n-1},a_{n-1}',b_{n-1-k},b'_{n-1-k}}W^{(R)}_{a_{n-1},a_{n-1}',b_{n-2},b'_{n-2}}\dots\\\nonumber
&\dots W^{\rm casc}_{b_{i+1},b'_{i+1},a_{i-k+2},a'_{i-k+2},a_{i+1},a_{i+1}',b_{i+1-k},b'_{i+1-k}} W^{\rm casc}_{b_{i},b'_{i},a_{i-k+1},a'_{i-k+1},a_i,a_i',b_{i-k},b'_{i-k}}\dots\\\nonumber
&\dots W^{(L)}_{b_1,b_1',a_1,a_1'}W^{\rm casc}_{b_{k},b'_{k},a_{1},a'_{1},a_k,a_k',b_{0},b'_{0}}W^{(L)}_{b_0,b_0',a_0,a_0'}\rho_{a_{0},a'_{0}}(t_0). 
\end{align}
This equation describes the tensor network in Fig.~\ref{fig:indexes}(a). This network consists of two types of transfer operators, $E^{[m]}(\Delta t)$ and $E^{[m-1]}(\Delta t)$, which we now explicitly define in terms of local propagators $W$ as
\begin{align}\label{eqn:EmW}
(E_i^{[p]}(\Delta t))_{a_{pk+i},\dots, a_i'}&\!=\!W^{(R)}_{a_{pk+i},a_{pk+i}',b_{pk+i-1},b'_{pk+i-1}}\!\!W^{\rm casc}_{b_{pk+i-1},b'_{pk+i-1},a_{pk+i-k},a'_{(p-1)k+i},a_{pk+i-1},a_{pk+i-1}',b_{(p-1)k+i-1},b'_{(p-1)k+i-1}}\!\!\dots\\\nonumber
&\dots W^{\rm casc}_{b_{i+k},b'_{i+k},a_{i+1},a'_{i+1},a_{i+k},a_{i+k}',b_{i},b'_{i}}W^{(L)}_{b_i,b_i',a_i,a_i'}, 
\end{align}
where $p=\{m,m-1\}$. Using this, we can write the expression in Eq.~\eqref{eqn:rhoWindex} using tensor network transfer operators
\begin{align}\label{eqn:rhoEindex}
\rho^{\rm sys}_{a_n,a'_{n}}(t_n)&=\sum_{{\bf a},{\bf b},{\bf a'},{\bf b'}}(E_{k-1}^{[m-1]}(\Delta t))_{a_{n},\dots, a_{k-1}'}\dots
(E_{i_r+1}^{[m-1]}(\Delta t))_{a_{mk+{i_r}+1},\dots, a_{i_r+1}'}(E_{i_r}^{[m]}(\Delta t))_{a_{mk+{i_r}},\dots, a_{i_r}'}\dots\\\nonumber
&\dots(E_1^{[m]}(\Delta t))_{a_{mk+1},\dots, a_1'} (E_0^{[m]}(\Delta t))_{a_{mk},\dots, a_0'}\rho_{a_{0},a'_{0}}(t_0), 
\end{align}
where $i_r= r/\Delta t$.
Thus, a calculation of the density matrix of the atom in front of the mirror results in performing the evolution of 1D cascaded chain.

The presence of the shifted periodic boundary conditions can be shown by considering the first and the last infinitesimal propagators $E_{k-1}^{[m-1]}(\Delta t)$ and $E_{0}^{[m]}(\Delta t)$ in the above expression. These propagators enter the sum with the number of coinciding indexes. To see this, we can compare the indexes of two arbitrary $W^{\rm casc}$ found at the same position in the definition~\eqref{eqn:EmW} of both propagators: $\{b_{(n+1)k-2},b'_{(n+1)k-2},a_{nk-1},a'_{nk-1},a_{(n+1)k-2},a_{(n+1)k-2}',b_{nk-2},b'_{nk-2}\}$ and $\{b_{nk-1},b'_{nk-1},a_{(n-1)k},a'_{(n-1)k},a_{nk-1},a_{nk-1}',b_{(n-1)k-1},b'_{(n-1)k-1}\}$ with $n\in\mathbb{N}$. Indexes $a_{nk-1},a'_{nk-1}$ coincide, and performing summation over these indexes leads to the contraction of the propagators $E_{k-1}^{[m-1]}(\Delta t)$ and $E_{0}^{[m]}(\Delta t)$, resulting in the shifted periodic boundary conditions (see also Fig.~\ref{fig:indexes}).
\end{widetext}

  \subsection{Infinite chain algorithm}\label{app:infchain}
  As discussed in the main text in the Sec.~\ref{sec:infchain}, the steady state of the atom in front of the mirror can be accessed by calculating the propagator of the infinite 1D cascaded chain. To do so, we make a translational invariant ansatz where all tensors are identical independently of the site they are associated with, $C^{[k]}(s)=C(s)$. We solve Eq.~\eqref{eq:Prop_differential} for this transitionally invariant infinite system size propagator using the infinite time-evolving block decimation algorithm (iTEBD) \cite{PhysRevLett.98.070201}. This is done using a two-site unit cell, with tensors denoted by $A(s)$ and $B(s)$ for even and odd sites, respectively. The integration of Eq.~\eqref{eq:Prop_differential} is achieved in a Trotterized fashion, where at each integration step we first apply $W_{\rm casc}$ to $A$ and $B$, then exchange the tensors and apply $W_{\rm casc}$ to $B$ and $A$. Note that this construction leads to tensors that are identical up to the Trotter errors, $A(\tau)=B(\tau)=C(\tau)$. 
  
\begin{figure}
             \includegraphics[width=0.8\linewidth]{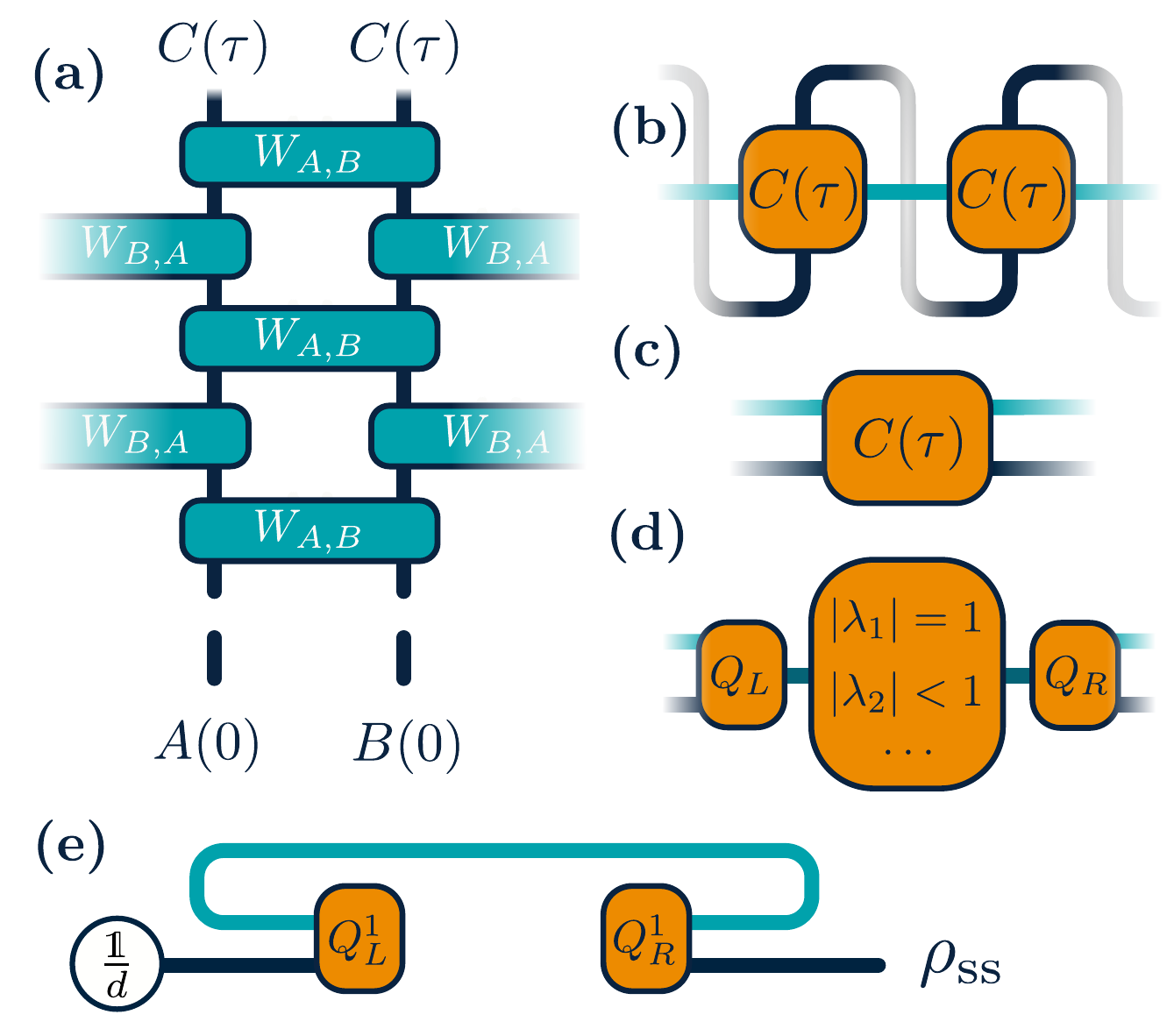}
				\caption{The infinite 1D chain algorithm: (a) Two tensors $A(0)$ and $B(0)$ are evolved until time $\tau$ using iTEBD by applying two-sites superoperators $W_{A,B}$ and $W_{B,A}$; (b) acquired after this evolution identical tensors $C(\tau)$ are to contracted together. To achieve that, we first rearrange the legs of the tensor $C(\tau)$ (c) and then apply spectral decomposition (d), where $Q_L$ and $Q_R$ are matrices consisting of the left and right eigenvectors. To calculate $\rho_{\rm ss}$, we contract an infinite number of the decomposed tensors; therefore, only vectors $Q^1_L$ and $Q^1_R$ associated with the eigenvalue $|\lambda_1|=1$ are left to be contracted (e).}
    \label{fig:infalgorithm1}
			\end{figure}

  Once the infinite system size propagator is obtained, the density matrix of  the atom in front of the mirror in the steady state, $\rho_{\rm ss}$, is obtained  by a contraction with shifted periodic boundary conditions, see Eq.~\eqref{eqn:rhoss}. To perform this contraction, we first reshape the tensor $C(\tau)$ such that it forms a square matrix of dimension $\chi d^2\times\chi d^2$ and subsequently calculate its eigenvalues $\lambda_\alpha$ as well as the matrices containing left and right eigenvectors, $Q_L$ and $Q_R$ (cf. Fig.~\ref{fig:infalgorithm1}(c)-(d)), i.e., $C(s)=\sum_\alpha (Q_L)_\alpha \lambda_\alpha (Q_R)_\alpha$. Introducing the diagonal matrix $\Lambda_{\alpha,\beta}=\delta_{\alpha,\beta}\lambda_{\alpha}$, the steady state is given by $\rho_{\rm ss}={\rm tr}_{\rm virt}\{\lim_{m\rightarrow\infty}C^m\}={\rm tr}_{\rm virt}\{\lim_{m\rightarrow\infty}Q_L\Lambda^mQ_R\}\mathbb{1}/d$, where we introduced ${\rm tr}_{\rm virt}\{\dots\}$ representing the trace over the virtual degrees of freedom (see Fig.~\ref{fig:infalgorithm1}(e)). Since $C(\tau)$ is a completely positive trace-preserving  map, its largest eigenvalue is of magnitude one, i.e.,  $|\lambda_1|=1$. If the steady state is unique, all other eigenvalues are smaller, i.e., $|\lambda_k|<1$ (for $k=2,3,\dots)$. Therefore, we can easily perform the total contraction in the infinite limit obtaining $\rho_{\rm ss}={\rm tr}_{\rm virt}\{\lim_{m\rightarrow\infty}Q_LD^mQ_R\}=Q^1_LQ^1_R\mathbb{1}/d$, where $Q^1_L$ and $Q^1_R$ are the left and the right eigenvector associated with the eigenvalue $|\lambda_1|=1$. 
  
  Another useful feature of the above procedure is that we can compute the time required to achieve the steady state $t_{\rm ss}$. Specifically, we can bound this time via the second largest eigenvalue of the transfer tensor $C(\tau)$, $|\lambda_2|^{\lceil t_{\rm ss}/\tau \rceil}=\exp(-t/t_{\rm ss})$, which describes how fast the information about the initial state fades with time (the number of sites in the chain). Thus we obtain the steady state time as
\begin{equation}
t_{\rm ss}=-2\frac{\tau}{\log_2{|\lambda_2|}}.
\end{equation}
   
\begin{figure}[!ht]
             \includegraphics[width=0.8\linewidth]{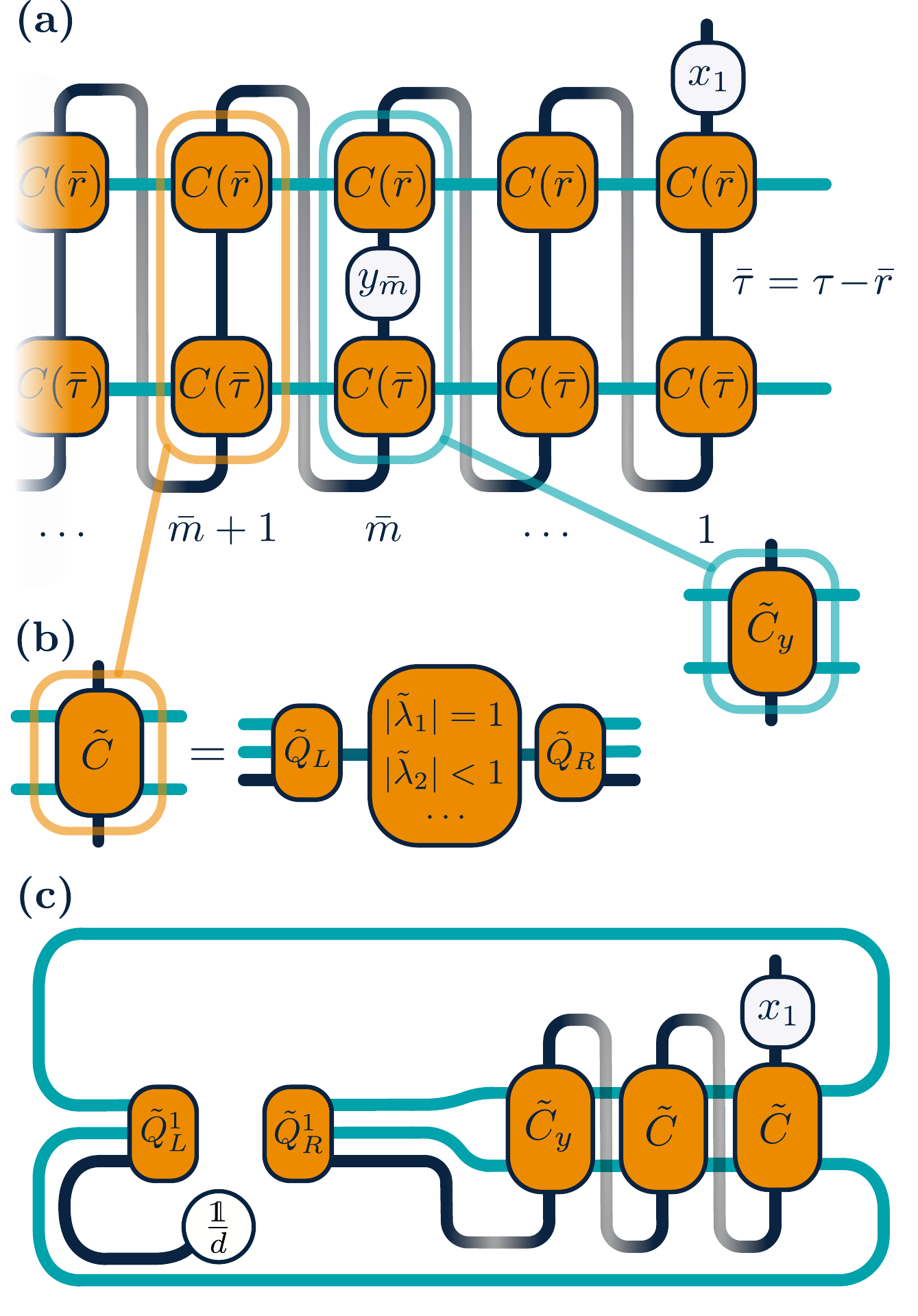}
				\caption{Steady state two-times correlation function algorithm: (a) To calculate the system correlation function, we first apply the system operators to two semi-infinite propagators with propagating times $\Bar{\tau}=\tau-\Bar{r}$ and $\Bar{r}$ and then contract the whole structure with the shifted periodic boundary conditions. (b) The contraction is performed using the spectral decomposition of a combined transfer operator $\Tilde{C}$ comprising $C(\Bar{r})$ and $C(\Bar{\tau})$ contracted over the physical leg. $\Tilde{C}$ is again a completely positive trace-preserving map with $|\Tilde{\lambda}_1|=1$. (c) We then contract an infinite number of $\Tilde{C}$ with $\rho=\frac{\mathbb{1}}{d}$ and the rest tensor structure on the right, and then trace over the virtual degrees of freedom. When taking the trace over the system degrees of freedom of the resulting structure, we obtain the correlation function $\lim_{t\rightarrow\infty}\langle x(t)y(t-t')\rangle$.}
    \label{fig:infcorfun}
			\end{figure}

Arbitrary system correlation functions as well as field correlation functions (using input-output formalism) can be calculated in the infinite limit in a similar way as in the case of the transient dynamics. Let us consider the example of two-times system correlation function $\lim_{t\rightarrow\infty}\left<x(t)y(t-t')\right>$. This expression depends only on the time difference $t'=\bar{m}\tau+\bar{r}$. Again we can write $\lim_{t\rightarrow\infty}\left<x (t)y(t-t')\right>={\rm tr}\{\mc{P}(M^{[\infty]})\frac{\mathbb{1}}{d}\}$ with $M^{[\infty]}$ defined as:
\begin{align}\label{eq:mt1inf}
    M^{[\infty]}=x_{1}E\strut^{[\infty]}(\bar{r})y_{\bar{m}} E\strut^{[\infty]}(\tau-\bar{r}),
\end{align}
where we use propagator $E\strut^{[\infty]}(t')$ defined as a propagator of 1D semi-infinite cascaded chain with infinitely many sites on the left. In contrary with the finite chain algorithm, here we count sites from the right (finite) side of the chain, thus $x_1$ denotes the operator $x$ acting on the rightmost replica in the chain and $y_{\bar{m}}$ acts on the replica located $\bar{m}$ sites away from the right as illustrated in Fig.~\ref{fig:infcorfun}. We perform the contraction of the infinite side of the chain again by means of the spectral decomposition. As for the transient case, the computational cost of calculating steady state $p$-times correlation function scales exponentially with $p$.

     \subsection{Multi-node setups generalization}\label{app:multiatom}
     We will now provide details on the generalization of the analysis described in the main text for the case of the multiple-node networks discussed in the main text. First, we consider the case of two nodes coupled to a bidirectional waveguide. Note that this setup can be equivalently interpreted as a network of two nodes (denoted $A$ and $B$) coupled to two unidirectional waveguides, where these waveguides represent the left- and the right-moving photons of the bidirectional waveguide, respectively. We also generalize this setting to the case of $n$ nodes interacting with $n$ unidirectional waveguides in setups of the form given in Fig.~\ref{fig:multinodes} for $n=3$. In each of these cases, time delays lead to an essential non-Markovianity due to the possibility of information to propagating in loops with time delays. Surprisingly, any multi-node problem with commensurate round-trip times between the nodes mediated by the unidirectional channel can be mapped to a set of Markovian 1D cascaded chains, e.g, the two-nodes problem maps to the evolution of two 1D cascaded chains, and consequently the three-nodes setup corresponds to the three 1D cascaded chains. 
     
     To start, let us consider the case of the two connected nodes shown in the Fig.~\ref{fig:Twoatoms}, for this configuration one can again build a tensor network representing the total wavefunction $\ket{\Psi(t)}$ of both nodes and the state of the waveguide. Similar to the single-node case, one can obtain the tensor network for the reduced density matrix of both nodes by tracing out the bath degrees of freedom $\rho_{\rm sys}(t)={\rm tr}_{ \mathcal{H}_B}\{\ket{\Psi(t)}\bra{\Psi(t)}\}$. The size of this network along the first  dimension is set by $2k$, i.e., by the round-trip time $2\tau$ in units of $\Delta t$, while the size along the second dimension is given by $m=\lceil n/k \rceil$, i.e., total evolution time $t_n-t_0$ in units of the time $\tau$, rounded up. Again we identify the transfer operator of the total network. For this network we find that there are two relevant transfer operators: These operators are the propagators describing the evolution of two 1D cascaded chains. The first chain consists of replicas of node A on odd sites and replicas of node B on even sites ($ABABAB\dots$), the second chain has an opposite order ($BABABA\dots$). We therefore call the first chain $AB$-chain, and the second chain $BA$-chain. Each chain has $m$ replicas, where $m$ is defined again through $t_n=m\tau+r$, with $0\leq r\leq\tau$ (see Fig.~\ref{fig:multinodes}(c)).
     The corresponding propagators for these chains satisfy the following  equations (analogous to Eq.~\eqref{eq:Prop_differential})
\begin{align}\label{eq:Prop_differential2}
\frac{d}{ds}E_{\{AB\}}^{[m]}(s)&=\mc{L}_{\{AB\}}^{[m]}E_{\{AB\}}^{[m]}\!(s), \\
\frac{d}{ds}E_{\{BA\}}^{[m]}(s)&=\mc{L}_{\{BA\}}^{[m]}E_{\{BA\}}^{[m]}\!(s),
\end{align}
with the Lindblad superoperators defined as
\begin{align}\label{eqn:bigLinbl2}
\mc{L}_{\{AB\}}^{[m]}&=\sum_{j\in {\rm odd}}\mc{L}^{\rm casc}_{A,B}+\sum_{j\in {\rm even}}\mc{L}^{\rm casc}_{B,A}+\mc{L}_{\{AB\}}^{\rm boundary},\\
\mc{L}_{\{BA\}}^{[m]}&=\sum_{j\in{\rm odd}}\mc{L}^{\rm casc}_{B,A}+\sum_{j\in {\rm even}}\mc{L}^{\rm casc}_{A,B}+\mc{L}_{\{BA\}}^{\rm boundary},
\end{align}
where the summation goes over odd (even) $j$ from $1$ to $m-1$ and $\mathcal{L}^{\rm casc}_{A,B}$ ($\mathcal{L}^{\rm casc}_{B,A}$) describes a cascaded coupling from replica $A$ to replica $B$ (from $B$ to $A$)
\begin{align}\label{eqn:cascL2AB}
\mathcal{L}^{\rm casc}_{A,B}X\!&=\!-\frac{i}{\hbar}\left[H_{A,B}^{\rm casc},X\right]+\mathcal{D}[R_A+L_{B}]X,
\end{align}
with the cascaded Hamiltonian
\begin{align*}
    H_{A,B}^{\rm casc}=\frac{1}{2}\lr{H_{\text{sys},A}+H_{\text{sys},B}+i\lr{ R_A^\dag L_{B}-L_{B}^\dag R_A}}
\end{align*}
and analogously \begin{align}\label{eqn:cascL2BA}
\mathcal{L}^{\rm casc}_{B,A}X\!&=\!-\frac{i}{\hbar}\left[H_{B,A}^{\rm casc},X\right]+\mathcal{D}[R_B+L_{A}]X,
\end{align}
with the cascaded Hamiltonian
\begin{align*}
    H_{B,A}^{\rm casc}=\frac{1}{2}\lr{H_{\text{sys},B}+H_{\text{sys},A}+i\lr{ R_B^\dag L_{A}-L_{A}^\dag R_B}}.
\end{align*}

The expression~\eqref{eqn:bigLinbl2} also contains boundary terms $\mc{L}_{\{AB\}}^{\rm boundary}$ and $\mc{L}_{\{BA\}}^{\rm boundary}$ acting on the first and the last replica of each chain, analogous to Eq.~\eqref{eqn:boundaryterms}.

To calculate the reduced density matrix of the nodes $A$ and $B$ at time $t_n$, we calculate the total propagators for the two chains, contract them with each other and with double-shifted periodic boundary conditions, and then apply the result to the initial density matrix (see Fig.~\ref{fig:multinodes}(d))
\begin{widetext}
\begin{align}\label{eqn:contr2nodes}
    \rho_{\rm sys}^{AB}(t_n)\!=\!\mathcal{P}_2(E^{[m-1]}_{\{{\rm AB}\}}(\tau-r)E^{[m]}_{\{{\rm AB}\}}(r)E^{[m-1]}_{\{{\rm BA}\}}(\tau-r)E^{[m]}_{\{{\rm BA}\}}(r))\rho^{AB}_{\rm sys}(t_0),
\end{align}
where we used $\mathcal{P}_2(X)$ to denote an application of the double-shifted periodic boundary conditions to the tensor network $X$. The cost of the total contraction in the Eq.~\eqref{eqn:contr2nodes} is $O(m \chi^{5}d^8)$. 

Note that one could also calculate one total propagator of one chain of the length $m+1$ ($ABABAB\dots$) by first applying the cascaded Lindblad propagators to the sites $2,\dots, m$ until time $r$, then propagating only sites $2,\dots, m-1$ until time $\tau$, followed by the evolution of the sites $1,\dots, m-1$ from time $r$ to time $\tau+r$, and finally propagating to time $2\tau$ by evolving the sites  $1,\dots, m-2$ (cf. Fig.~\ref{fig:multinodes}(d)). The resulting propagator is again contracted with the double-shifted periodic boundary conditions and the initial density matrix at a computational cost of $O(m \chi^{2}d^6)$. Even though this approach is more efficient for calculating the density matrix at a fixed time $t_n$, it requires iterating the entire calculation for each different time of interest. In contrast, the method described above allows one to compute the propagators for a fixed $\tau$ in parallel and then construct tensor networks for the various times of interest, tracing out the last replicas if needed during the process. This discussion applies to the calculation of the correlation functions described in the main text: One can evolve the cascaded chain and insert operators $x$ and $y$ at their right places during the evolution. The algorithm in the end must be chosen based on the specific task.

In the case of three nodes $A$, $B$, and $C$ connected in a loop via  unidirectional waveguides (see Fig.~\ref{fig:multinodes}(e)), the total tensor network for the reduced density matrix has three types of transfer operators. These operators are propagators for three cascaded chains consisting of replicas of the nodes: ($ABCABCABC\dots$), ($CABCABCAB\dots$), and ($BCABCABCA\dots$), as illustrated in Fig.~\ref{fig:multinodes}(e). The three resulting total propagators are contracted with the triple-shifted periodic boundary conditions and applied to the initial system density matrix $\rho_{ABC}(t_0)$ as depicted in Fig.~\ref{fig:multinodes}(f)
\begin{align*}
    \rho_{\rm sys}^{ABC}(t_n)=\mathcal{P}_3(E^{[m-1]}_{\{{\rm ABC}\}}(\tau-r)E^{[m]}_{\{{\rm ABC}\}}(r)E^{[m-1]}_{\{{\rm CAB}\}}(\tau-r)E^{[m]}_{\{{\rm CAB}\}}(r)E^{[m-1]}_{\{{\rm BCA}\}}(\tau-r)E^{[m]}_{\{{\rm BCA}\}}(r))\rho^{ABC}_{\rm sys}(t_0),
\end{align*}
where we denoted an application of the triple-shifted periodic boundary conditions to the tensor network $X$ as $\mathcal{P}_3(X)$. The contraction cost is $O(m \chi^{7}d^{10})$. 

The generalization to the larger number of the nodes $n$ $(A_1,\;A_2,\dots,\;A_n)$ in the setup thus requires the following steps. First, one needs to construct $n$ 1D cascaded chains. The unit cell of each chain is obtained using the cyclic permutation of the nodes order $(A_1,\;A_2,\dots,\;A_n)$. The second step is to calculate the total propagators for each chain with two-site superoperators, which are different for each chain as long as the nodes in the setup are not identical. This is followed by the contraction of the resulting $n$ propagators with the periodic boundary conditions shifted by $n$ sites, and applying the whole structure to the initial density matrix of the nodes. One can write a generalized expression for the system density matrix of $n$ nodes at time $t_n$ as
\begin{align*}
    \rho_{\rm sys}^{\{A_1,A_2,\dots,A_n\}}(t_n)=\mathcal{P}_n(E^{[m]}_{\{A_1,A_2,\dots,A_n\}}(\tau-r,r)E^{[m]}_{\{A_n,A_1,\dots,A_{n-1}\}}(\tau-r,r)\dots E^{[m]}_{\{A_2,A_3,\dots,A_1\}}(\tau-r,r))\rho^{\{A_1,A_2,\dots,A_n\}}_{\rm sys}(t_0),
\end{align*}
where we used the shorthand notation $E^{[m]}(\tau-r,r)=E^{[m-1]}(\tau-r)E^{[m]}(r)$. While the chain evolution can be performed in parallel, the cost of propagators' contraction scales exponentially with the number of nodes $O(m \chi^{3+n}d^{4+2n})$. 
\end{widetext}
   \bibliographystyle{apsrev4-2-mod}
\bibliography{Library.bib, HPbib.bib}

\end{document}